\RequirePackage{fix-cm}
\documentclass[natbib,smallextended]{svjour3}       % onecolumn (second format)
\smartqed  % flush right qed marks, e.g. at end of proof
\usepackage{graphicx}
\usepackage{enumerate}

\usepackage[usenames,dvipsnames]{color}
\usepackage{ulem}  %%%% for strikeout font (\sout)   %%%%% to indicate suggestions for deletions in the text
 \journalname{my journal}
\newcommand{\bdot}{\mbox{\boldmath $\cdot$}}
\newcommand{\del}{\mbox{\boldmath $\nabla$}}
\newcommand{\cross}{\mbox{\boldmath $\times$}}

\newcommand{\vv}{\mbox{\boldmath $v$} {}}
\newcommand{\bb}{\mbox{\boldmath $b$} {}}
\newcommand{\jj}{\mbox{\boldmath $j$} {}}
\newcommand{\nab}{\mbox{\boldmath $\nabla$} {}}
\newcommand{\aaaa}{\mbox{\boldmath $a$} {}}
\newcommand{\EQ}{\begin{equation}}
\newcommand{\EN}{\end{equation}}
\newcommand{\uu}{\mbox{\boldmath $u$} {}}
\newcommand{\meanBB}{\overline{\mbox{\boldmath $B$}}{}}{}
\newcommand{\ee}{\mbox{\boldmath $e$} {}}
\newcommand{\Eq}[1]{Equation~(\ref{#1})}

\begin{document}

\title{Large-Eddy Simulations of Magnetohydrodynamic Turbulence in Heliophysics and Astrophysics}


\titlerunning{LES of MHD Turbulence}        % if too long for running head

\author{Mark Miesch \and
        William Matthaeus \and
        Axel Brandenburg \and
        Arakel Petrosyan \and
        Annick Pouquet \and
        Claude Cambon \and
        Frank Jenko \and
        Dmitri Uzdensky \and
        James Stone \and
        Steve Tobias \and
        Juri Toomre \and
        Marco Velli
}


\newpage
\institute{M. S. Miesch \at
              HAO/NCAR, 3080 Center Green Dr.\ Boulder, Colorado, 80301, USA \\
              Tel.: +303-497-1582, Fax: +303-497-1589 \\
              \email{miesch@ucar.edu}           %  \\
           \and
           W. H. Matthaeus \at
           Dept.\ of Physics and Astronomy, University of Delaware \\
           Newark, DE, 19716, USA \\
           whm@udel.edu
           \and
           A. Brandenburg \at
           NORDITA, KTH Royal Inst.\ Tech.\ and Stockholm Univ. \\
           Roslagstullsbacken 23, SE-10691, Stockholm, Sweden \\
           brandenb@nordita.org
           \and
           A. Petrosyan \at
             Space Research Inst., Russian Academy of Sciences,
             Profsoyuznaya 84/32, Moscow, Russia \\
             Moscow Institute of Physics and Technology (State University), 
             9 Institutskiy per., Dolgoprudny, Moscow Region, 141700, Russia \\
             \email{a.petrosy@iki.rssi.ru}
           \and
           A. Pouquet \at
           NCAR, P.O. Box 3000, Boulder, CO, 80307, USA \\
           and Dept.\ of Applied Math., Univ. of Colorado \\
           Boulder, CO, 80309-256, USA \\
           pouquet@ucar.edu
           \and
           C. Cambon \at
           Laboratoire de M\'ecanique des Fluides et d'Acoustique \\
           \'Ecole Centrale de Lyon \\
           36, av., Guy de Collongues, F-69134, Ecully cedex, France \\
           claude.cambon@ec-lyon.fr
           \and
           F. Jenko \at
           UCLA, Dept.\ of Physics and Astronomy \\
           475 Portola Plaza, Office 4-720 PAB, Box 951547 \\
           Los Angeles, CA 90095-1547, USA \\
           jenko@physics.ucla.edu
           \and
           D. Uzdensky \at
           CIPS, Physics Dept., UCB-390, Univ. Colorado \\
           Boulder, CO 80309-0390, USA \\
           Uzdensky@colorado.edu
           \and
           J. Stone \at
           Dept.\ of Astrophysical Sciences, Princeton Univ.\ \\
           Peyton Hall, Ivy Lane, Princeton, NJ 08544-1001, USA \\
           jstone@astro.princeton.edu
           \and
           S. Tobias \at
           Dept.\ of Applied Mathematics, University of Leeds \\
           Leeds, LS2 9JT, UK 
           \and
           J. Toomre \at
           JILA and Dept.\ of Astrophysical and Planetary Sciences \\
           Univ.\ of Colorado, Boulder, CO, 80309-0440, USA \\
           jtoomre@lcd.colorado.edu
           \and
           M. Velli \at
           Jet Propulsion Laboratory, California Inst.\ Tech. \\
           Pasadena, CA, 91109-8099, USA \\
           mvelli@jpl.nasa.gov
}

\date{Received: date / Accepted: date}

\maketitle

\begin{abstract}
We live in an age in which high-performance computing is transforming 
the way we do science.  Previously intractable problems are now becoming
accessible by means of increasingly realistic numerical simulations.  
One of the most enduring and most challenging of these problems is turbulence.  
Yet, despite these advances, the extreme parameter regimes encountered in 
space physics and astrophysics (as in atmospheric and oceanic physics)
still preclude direct numerical simulation.  
Numerical models must take a Large Eddy Simulation (LES) approach, explicitly 
computing only a fraction of the active dynamical scales.  The success of
such an approach hinges on how well the model can represent the subgrid-scales
(SGS) that are not explicitly resolved.  In addition to the parameter regime, 
heliophysical and astrophysical applications must also face an equally daunting 
challenge: magnetism.   The presence of magnetic fields in a turbulent, 
electrically conducting fluid flow can dramatically alter the coupling
between large and small scales, with potentially profound implications 
for LES/SGS modeling.  In this review article, we summarize the state 
of the art in LES modeling of turbulent magnetohydrodynamic (MHD) flows.
After discussing the nature of MHD turbulence and the small-scale processes
that give rise to energy dissipation, plasma heating, and magnetic
reconnection, we consider how these processes may best be captured 
within an LES/SGS framework.  We then consider several specific applications
in heliophysics and astrophysics, assessing triumphs, challenges, and 
future directions.
\end{abstract}

%========================================================================
\section{Introduction}
\label{sec:intro}

%{\it {
%\AGP{My changes are in magenta; remark that small changes are not indicated}.
%\ARP{ ARP, in green, could be for Arakel;}
%\PS{PS, in blue, for Pierre, and}
%\MSM{MM, in red, for Mark.}
%\AGP{This is just a suggestion. Note that I also indicate TEXT that I suggest to be suppressed with the backslash ``sout'' \sout{TEXT} command.}
%}}

% MSM

On May 20--23, 2013 a workshop was held at the National Center for Atmospheric Research (NCAR) in 
Boulder, Colorado, USA entitled ``Large-Eddy Simulations (LES) of Magnetohydrodynamic (MHD) Turbulence.''
The workshop was sponsored by NCAR's Geophysical Turbulence Program (GTP) and involved approximately
fifty participants from eight countries.

This review paper is a product of the GTP workshop, though it is not intended as a comprehensive 
account of the proceedings.  Rather, it is intended as a summary of the issues addressed and 
the insights achieved, as well as an inspiration and a guide to promote future work on this subject.
Though the subject of interest, namely LES of MHD turbulence, is ostensibly rather specific,
it encompasses a number of subtle physical processes and diverse applications and it draws on the 
formidable discipline of efficient numerical algorithm development on high-performance computing 
architectures.

The fundamental challenge that defines the field of LES is that the range of dynamical scales active in 
many turbulent fluid systems far exceeds the range that can be explicitly captured in a 
computer simulation.   Examples include the convection zones of stars, planetary atmospheres, astrophysical 
accretion disks, and industrial applications such as gas turbines.  The {\it central premise} of LES is that
large scales dominate the turbulent transport and energy budget so a numerical simulation that captures
those scales explicitly will provide a realistic depiction of the flow for all practical purposes, provided 
that the small scales that cannot be resolved are somehow taken into account.  Strategies for incorporating
the small scales include explicit subgrid-scale (SGS) models or implicit numerical dissipation 
schemes.  

\begin{figure}[t!]\begin{center}
\includegraphics[bb=0.0in 0.0in 11.0in 11.0in,width=.98\textwidth]{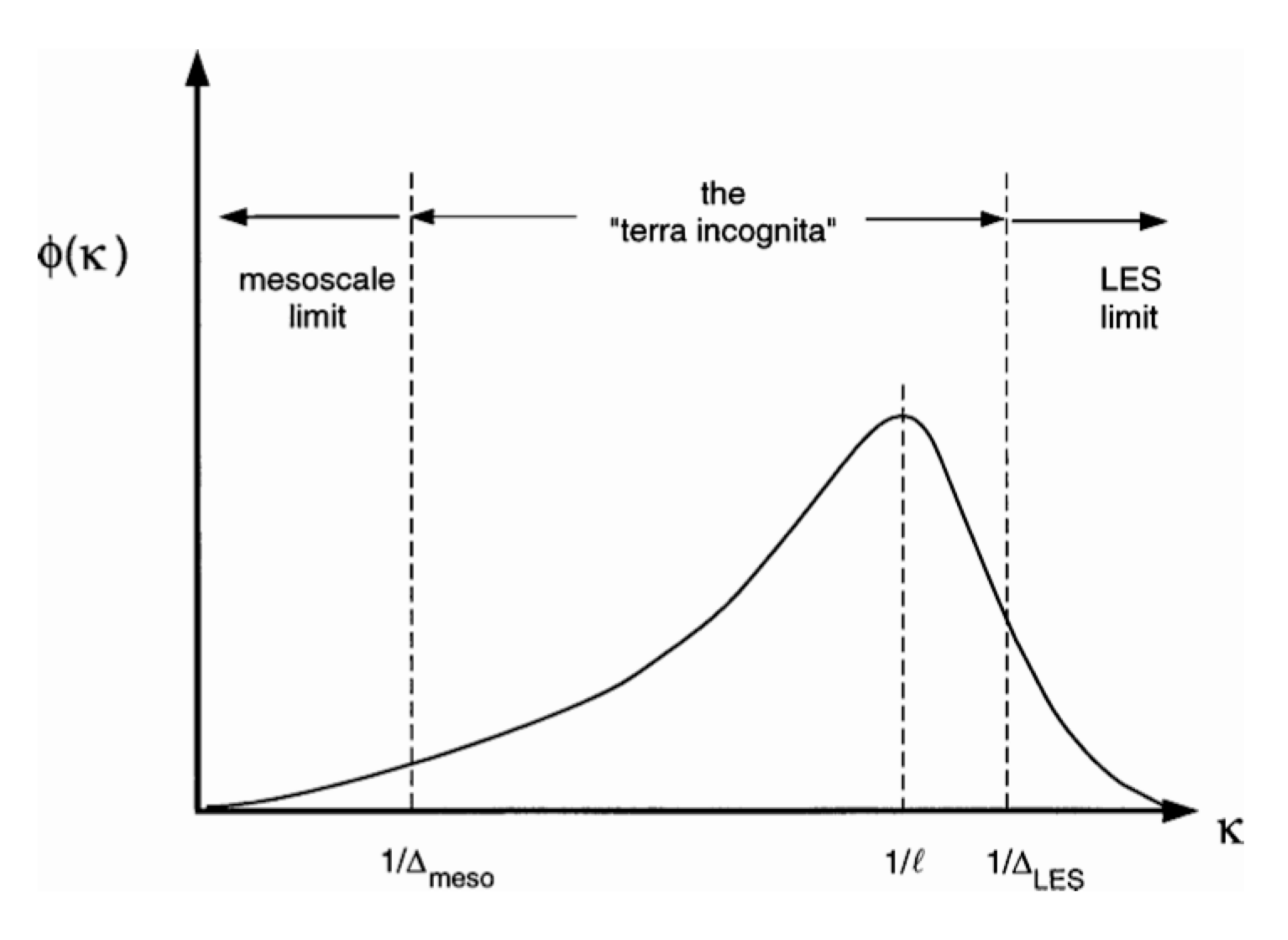}
\end{center}
\caption{The ``Terra Incognita'' and the realm of LES \citep[from][presented at the workshop by Peter Sullivan]{wynga04}.  Shown is an idealized
  power spectrum of some turbulent field $\phi$ as a function of wavenumber $\kappa$ (referred to elsewhere
in the paper as $k$).  The peak of the spectrum lies at $\kappa \sim 1/\ell$ where $\ell$ is a characteristic 
length scale of the turbulence.  If the grid spacing of the simulation, $\Delta$ is much larger
than $\ell$, then the turbulence is entirely unresolved, defining the so-called mesoscale
limit of atmospheric science ($\Delta \geq \Delta_{\mbox{meso}}$).  If the turbulence is partially
resolved, capturing the peak in the spectrum but not the viscous dissipation scale $\ell_d$ 
($\ell_d \leq \Delta \leq \ell$), then this is the appropriate scenario for LES.\label{fig:incognita}}
\end{figure}

The range of validity for LES is illustrated schematically in Fig.\ \ref{fig:incognita}.  Consider
a numerical simulation of a turbulent fluid system in which the turbulent energy spectrum peaks at
some characteristic wavenumber $\ell^{-1}$.  Due to the nature of digital computing, any such simulation
can only capture a finite range in wavenumber, say from $L^{-1}$ to $\Delta^{-1}$.  Guided by the 
{\it central premise} stated above, the lower bound of this wavenumber range often corresponds to 
the largest scales in the system $L > \ell$.  Meanwhile, the higher bound in wavenumber is determined 
by the resolution limit $\Delta$ which may correspond to a numerical grid spacing or to the effective 
width of some explicit low-pass filtering operation that averages over the smaller scales (\S\ref{sec:lesmhd}).

If the resolution limit $\Delta$ is smaller than the viscous, thermal, and magnetic dissipation scales, 
collectively represented here as $\ell_{\rm diss}$, then the simulation may be regarded as a direct numerical 
simulation (DNS).  However, as noted above, DNS are not possible for most turbulent 
systems in astrophysics and space physics.  A much more tractable situation is when the resolution 
limit captures the turbulence scale $\ell$ but not the dissipation scales; $\ell_{\rm diss} \ll \Delta \ll \ell$.
This is the realm of LES.  Here SGS models can exploit the self-similarity of the turbulent cascade
in the inertial range and diffusive prescriptions are often sufficient (although even that may not
be true in MHD).   Ideally, the grid spacing $\Delta$ should also be much less than other scales that 
lead to large-scale anisotropy, such as the Rossby deformation radius, the Bolgiano scale of convection, 
and the pressure and density scale heights.  When this is not possible, such sources of anisotropy must 
be taken into account in any explicit SGS model.

Sometimes the characteristic scale of a turbulent flow component is smaller than the effective 
resolution of the simulation or model $\ell < \Delta$.  
One may then model the influence of the unresolved scales on mean, resolved flows 
but one might not refer to this as an LES model.  
A better terminology might be to call this a Reynolds-Averaged Navier-Stokes (RANS) approach with 
some model for the turbulent transport, possibly including non-diffusive as well as diffusive 
components.  In the following, such calculations are also referred to as mean-field simulations (MFS).

If properly formulated, the LES approach should converge to the DNS approach as $\Delta$ goes to 
zero.  This is not necessarily true for RANS.  For many systems such as homogeneous turbulence, 
there is a smooth transition from RANS to LES as the filter scale is decreased from 
$\Delta > \ell$ to $\Delta \ll \ell$ \citep{schmi15}.  SGS models may include non-diffusive 
transport that resembles the Reynolds stress modeling in a RANS system, blurring the distinction 
between the two approaches.  Numerical models of such systems may lie anywhere along a continuous
spectrum of modeling approaches from RANS to LES to DNS.  On the RANS end 
of the spectrum, the reliability of the Reynolds stress model is paramount, along with 
analogous prescriptions for turbulent heat transport and, in the MHD case, turbulent magnetic induction.  
For LES models that lie more toward the DNS end of the spectrum the details of the SGS model 
presumably become less important, though the simulation becomes more sensitive to the accuracy of the
numerical algorithm.  Also, as one moves across the spectrum from RANS to LES to DNS the
computational cost increases greatly, along with the number of degrees of freedom.

In other systems, the transition from RANS to LES is less straightforward; one must beware of the
``Terra Incognita'' that may lie between (Fig.\ 1).  As the LES filter size $\Delta$ approaches 
$\ell$, SGS models that rely on the self-similar nature of turbulent cascades may break down.
There may be a maximum scale $\Delta_{LES} < \ell$ above which the filtering procedure
becomes ill defined and unreliable.  On the other hand, RANS models may require a sufficient
scale separation to make statistical averages meaningful, such that $\Delta \gg \ell$.
In Fig.\ 1 this minimum scale for the validity of RANS is labelled as $\Delta_{\rm meso}$,
in reference to the mesoscale modeling of the Earth's atmosphere. In between these two limits, 
$\Delta_{\rm LES} < \Delta < \Delta_{\rm meso}$, lies the Terra Incognita where turbulence modeling 
and simulation can become much more challenging \citep{wynga04}.

Due largely to the industrial and atmospheric applications, LES of hydrodynamic turbulence is 
widespread and relatively mature \citep{sagaut06}.  However, most astrophysical and geophysical flows of 
interest are electrically conducting plasmas in which the magnetic field plays an essential 
dynamical role.  For these flows, models must take magnetism into account either through the 
kinetic theory of plasmas (generally necessary for the smallest scales) or through the simplifying
equations of MHD (often well justified for large scales).   Though LES of MHD turbulence can build upon 
the large body of work in hydrodynamic (HD) turbulence, it poses unique challenges that must be addressed specifically.
These include small-scale anisotropy, nonlocal spectral transfer, and magnetic reconnection.

In section 2 we discuss some of these unique challenges of MHD turbulence and highlight particular 
features of MHD turbulence that may promote the development of reliable SGS models.  In section 3 
we consider the physics of the smallest scales where ideal MHD no longer applies, promoting mechanical 
and magnetic energy dissipation and magnetic reconnection, and we ask how these scales may influence 
the dynamics of the large scales.  We then review current SGS modeling approaches for MHD in section 4 
and assess the triumphs and tribulations of current applications in section 5.  We summarize the 
state of the field in section 6 and anticipate where it may be headed in the future.

Though many of the physical processes and challenges we address have implications throughout astrophysics, we will focus primarily on solar and space physics in this review.  This is done to allow us to achieve some depth in the material covered while still maintaining a manageable length.  For a more comprehensive overview of LES/SGS in astrophysics the reader is referred to \cite{schmi15}.

\section{MHD Turbulence: Challenges and Building Blocks}
\label{sec:large}

%\subsection{from Steve Tobias}

\subsection{Anisotropy in incompressible  unbounded turbulence, from HD to MHD}

Global anisotropy is an essential feature of MHD flows, particularly
in the presence of a mean magnetic field.  The existence of a unique fixed 
orientation yields breaking isotropy towards axisymmetry, with or without 
mirror symmetry.   System rotation, buoyancy, and density stratification
further contribute to global anisotropy and inhomogeneity as in HD turbulence.  
However, in these HD cases, if one considers scales small enough (e.g. much
smaller than the Rossby
radius of deformation for rotation) then the constraints become
negligible. For MHD turbulence the situation is exactly the
opposite; the flow never "forgets" the existence of the large-scale
constraint imposed by the magnetic field. Indeed as one goes to
smaller and smaller scales the anisotropy increases \citep{tobia13}.
This is a severe constraint that must be respected by sub-grid 
scale models.

In the absence of a mean magnetic field, Alfv\'enic MHD turbulence can 
be investigated with a more \citep{Iroshnikov} or less \citep{Kraichnan} 
isotropized model. However, even in this case, the substructure of 
Alfv\'en wave packets at small scales cannot be ignored in the overall 
structure and dynamics of the turbulence.  If the governing 
orientation of the small-scale Alfv\'en packets is seen as random, 
a sophisticated stochastic model, mixing anisotropy and intermittency, 
is needed (one such model was discussed by W.\ Matthaeus at the workshop).  
Again, this is a formidable challenge facing SGS in MHD.

The theory for MHD turbulence has been developed over the past few
decades, and there are many recent reviews summarizing various
aspects 
\citep{KraichnanMontgomery80, biskamp03, ZhouEA04,petro10,BN11,tobia13}.
One familiar phenomenology is that of interacting
wavepackets. This phenomenology arises because nonlinear Alfv\'en
waves are exact solutions of the full incompressible MHD
equations \citep[see, e.g.][]{Parker}.
A more precise statement is that nonlinear interactions
only take place when oppositely-signed Els\"asser fields
${\bf Z}_+ = \vv + \bb$ and ${\bf Z}_- = \vv-\bb$ overlap in space,
this statement being valid with or without a mean magnetic field
${\bf B}_0$, and even in two-dimensional geometry with ${\bf B}_0 = 0$
where there is no global propagation direction at all. 
In any case one often encounters the heuristic explanation
that interactions only take
place when oppositely propagating wave-packets interact with each
other. 
When coherent propagation can occur, it anisotropically 
interferes with nonlinearity,  and gives rise to
anisotropic spectra 
\citep{ShebalinEA83,OughtonEA94}. 
%%\MSM{due to the dynamic alignment of ${\bf v}$ and ${\bf B}$}. 
%% WHM eliminated the above phrase, as I believe it is not correct. We can discuss this!!
A related effect, the dynamic alignment 
of turbulent velocity and magnetic fields,
also has strong effects on MHD turbulence. 
Global dynamic alignment may occur in some ranges of parameter 
space
\citep{DobrowolnyEA80,TingEA86,StriblingMatthaeus91,StawarzEA12}
as a form of long time turbulent 
relaxation. However local dynamic alignment 
\citep{MilanoEA01,Boldyrev06,MatthaeusEA08}
occurs rapidly in turbulence.
Other types of local relaxation that reduce or suppress the strength of nonlinearities
imply formation of local patches of correlation 
associated with Beltrami velocity 
fields and force-free magnetic fields
\citep{ServidioEA08}.
Numerical experiments also seem to
indicate that the degree of alignment of field and velocity is
scale-dependent, with the alignment variation even propagating into
the dissipative regime \citep{Boldyrev06,MasonEA06}. 

For MHD turbulence with a 
strong externally supported DC magnetic field ${\bf B}_0$,
it is possible to form a large-scale condensate of energy
which influences 
the turbulent cascade at all smaller scales \citep{DmitrukMatthaeus09}. 
Condensation, whether of this
type, or of the inverse cascade type,
may be associated with generation of low frequency $1/f$ noise 
and long time correlations and sporadic level changes of 
energy and other quantities over very long times \cite{DmitrukMatthaeus07}.
 All of these dynamical
effects may influence computed solutions, and 
should be respected by appropriate sub-grid scale models.
These factors,
which present  a formidable challenge for SGS prescriptions, are discussed
in more detail in section \ref{sec:small}.

%CC
It is possible to describe the second-order correlation tensors with a 
minimal number of correlators, as scalar or pseudo-scalar spectra, 
accounting for the solenoidal properties of both velocity and vorticity
fields. The seminal studies by \citet{rober40}, \citet{chand50,chand51},
\citet{batch82}, and \citet{craya58} were completed by
\citet{oughton97} in the MHD case. 
Developed independently by Cambon's team, a similar formalism improved the
decomposition in terms of energy, helicity and especially  {\it polarization}
spectra, using the orthonormal bases for solenoidal fields, known as
a Craya-Herring frame of reference \citep{herri74}, with its variant of  
helical modes \citep{cambo89,waleffe92}.
This formalism is discussed at length, with application to turbulence
subjected to rotation, density-stratification and uniform shear in the 
recent monograph by \citet{sagaut08}, and extended to the  MHD
case by Cambon and collaborators \citep{favier12,cambon12}.
In addition to the definition of
the basic set of spectra and co-spectra, dynamical 
equations can be written for the correlators,
generalizing the Lin equation in isotropic turbulence. 

Unfortunately, very few of these results (for both HD and MHD) have been used 
in recent pseudo-spectral DNS in triple-periodic boxes, even if they could reproduce
anisotropic homogeneous turbulence, despite the finite-box effects, standard 
discretization,  questionable ergodicity from a single realization, and other 
differences with the theoretical context of homogeneous unbounded turbulence.
As a first example, helicity cannot be disentangled from directional 
anisotropy (e.g. angle-dependent, or two-component, energy spectrum) and
polarization anisotropy in DNS
started with a single realization, e.g. with ABC artificial helical forcing
\citep{salhi14}. 
On the other hand, angle-dependent spectra and co-spectra, which are not
provided by these recent DNS, are useful to {\it quantitatively} characterize different 
anisotropic properties, as the horizontal layering in stably-stratified 
turbulence, and the opposite trend to generate columnar structures in
flows dominated by system rotation. Such structures are often shown only  on snapshots in recent DNS, with  very indirect linkage to statistical indicators, such
as one-component, in terms of wavevector modulus $k$ or transverse wavevector
$k_{\perp}$, spectra.  Might there be some analogy between the layering in stably 
stratified turbulence (which is linked to the kinetic energy cascade of the 
toroidal velocity component and angle-dependent spectra) and the formation 
of thin current sheets in MHD, as seen in high-resolution DNS?

In addition, the distinction between the 2D `vortex' modes and 'rapid'
inertial modes is dependent on discretization in conventional
pseudo-spectral DNS for purely rotating turbulence, and the dynamics
is affected by finite-box effects. Only the use of actual confinement,
as with rigid boundaries, allows one to identify the 2D mode as a dominant
one, whereas it is only a marginal limit of inertial wave modes in a
very large box, and treated as
an integrable singularity in wave turbulence theory \citep{bellet06}. Extension of
inertial wave turbulence theory, with coupling to 'actual' 2D modes, was
recently achieved in a rotating 'slab' by Scott (2014).

An important question, useful for SGS modeling in LES, is the range of penetration of anisotropy 
towards smallest scales (see also \S\ref{sec:ssc}).  In the HD case, an external effect such  as mean
shear firstly affects the largest scales, generating both energy (production)
and anisotropy. As suggested by \cite{corrsin58}, isotropy can be recovered at a
typical wavenumber, expressed in terms of mean shear rate $S$ and the 
dissipation rate $\varepsilon$: $k_S = \sqrt{S^3 / \varepsilon}$. Similar threshold 
wavenumbers were proposed by \cite{ozmidov65} for stably-stratified turbulence,
replacing $S$ by the Brunt-V\"ais\"al\"a frequency, $N$, and by \cite{zeman94} in
rotating turbulence, replacing $S$ by system vorticity. Even if these 
simple dimensional considerations are only partly supported by DNS or experiments
\citep{lamriben11,delache14},
they are not sufficient to close the problem and to say that anisotropy 
can be generally neglected at small scale in HD turbulence, in contrast
with MHD turbulence. Rotating turbulence and stably stratified turbulence
are much more subtle, because there is no direct production of kinetic 
energy by the Coriolis force, and no direct production of total energy,
kinetic + potential, by the buoyancy force with stabilizing mean density
gradient (in contrast with turbulence subjected to mean shear). 
On the other hand, a scale-by-scale analysis of the anisotropy
in rotating turbulence, without artificial forcing, reveals that the anisotropy
first increases with increasing wave-number, so that it can be maximum
at the smallest scales if the ``Zeman wavenumber'' $k_{\Omega} = \sqrt{\Omega^3 / \varepsilon}$ 
is larger than the viscous cutoff.
These considerations suggest a refined comparison between inertial wave 
turbulence theory and weak MHD Alfv\'enic turbulence, with the latter
reviewed and updated by Boldyrev in the GTP workshop \cite[see][]{tobia13}. 

%A final point, addressed by Bill Matthaeus, emphasizes the importance
%of local anisotropy in inhomogeneous turbulence and its possible linkage
%to intermittency.
%is beyond the case of homogeneous 
%turbulence, and stress the importance of 
%Local anisotropy, with its possible linkage to intermittency. 

%AB:
\subsection{Is there a need for including advanced backscatter modeling?}
\label{sec:backscatter}

In HD turbulence, backscatter to larger scales plays energetically a
significant role, but it is usually not systematically correlated with 
large-scale properties of the flow.
On the other hand, at least in helical MHD, backscatter plays a dramatic
role in that it is responsible for the generation of magnetic energy at the
largest scale through what is known as the $\alpha$ effect.
The $\alpha$ effect plays therefore an important role in mean-field
simulations (MFS), but is ignored in LES.

The $\alpha$-effect is linked to the upscale transfer of magnetic helicity,
which occurs in helical MHD turbulence through local (inverse cascade) 
or nonlocal ($\alpha$-effect) spectral interactions
(Pouquet et al 1976; Seehafer 1996; Brandenburg 2001; M\"uller et al.\ 2012).  Spectral 
transfer of cross helicity $\left<\mbox{\boldmath $u \cdot B$}\right>$
can also couple large and small scales and should be taken into account in 
SGS models, as emphasized in the GTP workshop by Yokoi (2013).

\begin{figure}[t!]\begin{center}
\includegraphics[width=.98\textwidth]{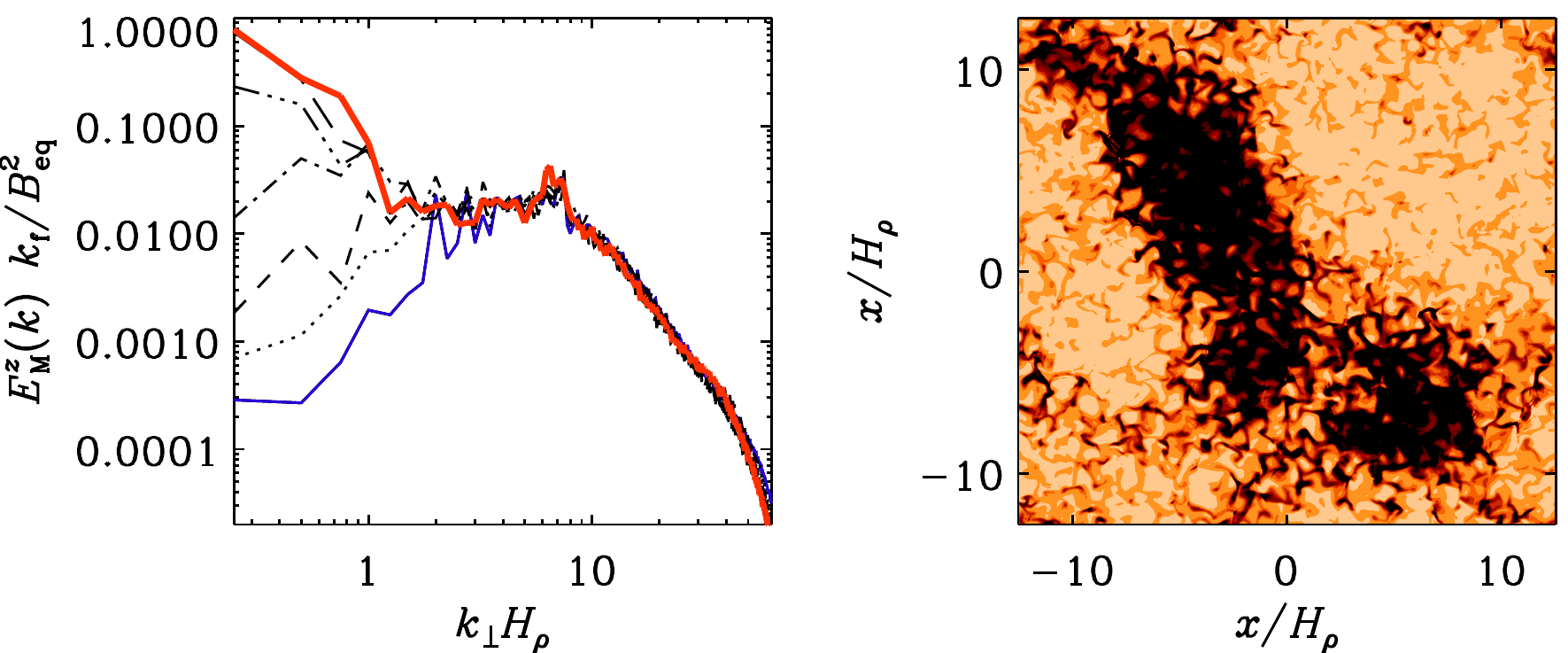}
\end{center}\caption[]{
Left: Normalized energy spectra of $B_z$ from an isothermally stratified,
randomly forced DNS with $g/c_{\rm s}^2k_{\rm f}=4$ (sound speed $c_{\rm s}$ and forcing
wavenumber $k_{\rm f}$) at turbulent diffusive times $\approx 0.2$ (blue line),
0.5, 1, 2, 5, 10, and 20 (red line).
Right: Magnetic field configuration at the upper surface near the end
of the simulation.
Adapted from \cite{BGJKR14}.
}\label{pBz_spec2__pBzm_top_comp_gkf}\end{figure}

Indeed, cross helicity is produced in the presence of gravity $\vec{g}$
and a parallel magnetic field $\vec{B}$, giving rise to a pseudo-scalar
$\vec{g\cdot B}$ that is odd in the magnetic field, just like the
cross helicity \citep{RKB11}.  In such a case, a large-scale magnetic pattern 
emerges, as can be seen from power spectra and images shown in
Fig.~\ref{pBz_spec2__pBzm_top_comp_gkf}.
Whether or not this large-scale pattern is a result of some inverse
cascading of cross helicity, analogous to the $\alpha$-effect,
remains an open question; see \cite{BGJKR14} for details.

Another potential mechanism that may contribute the the generation
of large-scale structure in MHD flows such as that shown in
Fig.~\ref{pBz_spec2__pBzm_top_comp_gkf} is the suppression of
small-scale turbulent pressure by a large-scale magnetic field.  This is currently gaining attention
within the context of mean-field modeling of the Reynolds and
Maxwell stresses.  If this suppression dominates over the direct
contribution of the magnetic pressure, as is the case for fully developed
turbulence, then the net effect will be negative \citep{Kleeorin,Rogachevskii}.
In a strongly stratified layer, this can lead to an instability, which
is now called the negative effective magnetic pressure instability (NEMPI).
It has been suggested that NEMPI may play a role in causing the
magnetic field to form flux concentrations in the upper regions of the
solar convection zone, where the stratification is 
strongest (Brandenburg et al.\ 2012; K\"apyl\"a et al.\ 2012; 
Kemel et al.\ 2013).
Again, this effect has been successfully captured in MFS, where its
predictive capabilities have been instrumental in furthering our theoretical
understanding.
We need to ask whether or not LES should be modified to include this effect,
or whether LES are naturally able to capture this type of physics.
For example, the dynamic Smagorinsky model used in simulations of stellar
convective dynamos by \cite{Nelson11,Nelson13} promotes the generation
of coherent flux structures by nonlinear feedbacks that are roughly analogous 
to those responsible for the NEMPI.

%AB:
\subsection{Possible consequences of misrepresenting the small scales}
\label{sec:misrepresenting}

A practical goal of LES is clearly to keep the code stable.
This means that close to the grid scale the flow must become smooth.
In reality, the opposite is the case: turbulent diffusion of mean flows,
mean magnetic fields, mean temperature, and mean passive scalars decreases
with scales.
This picture is quite clear in mean-field theory, where turbulent transport
coefficients such as magnetic diffusivity, $\eta_t$, and $\alpha$ effect are
known to become wavenumber-dependent.
We do not yet know whether this plays an important role in LES,
but we must ask whether certain discrepancies between LES and astrophysical
reality can be explained by such shortcomings.
Below we discuss one such example.

Realistic global dynamo simulations have revealed that magnetic cycles
are possible at rotation rates somewhat faster than the Sun \citep{BMBBT11}.
Yet, the Sun is known to undergo cycles.
Could this be a consequence of misrepresenting the small scales in the
simulations?

To approach this question, we need to know what is the governing non-dimensional
parameter that determines the transition from cyclic to non-cyclic dynamos.
This is a difficult question, because the mechanism behind the solar dynamo
is not conclusively identified.
Broadly speaking, there are flux transport dynamos where meridional circulation
plays an important role determining the cycle time and migration of magnetic
activity belts.
The other candidate is just $\alpha$ effect and differential rotation,
giving rise to an $\alpha$--$\Omega$ dynamo in which meridional circulation
is unimportant.
In mean-field theory, the relative importance of the $\Omega$ effect
for both scenarios is determined by the non-dimensional quantity
$C_\Omega=\Delta\Omega/\eta_t k^2$.  Here $\eta_t$ is a measure of the
turbulent kinetic energy so $C_\Omega$ may be regarded as equivalent to
a Rossby number based on the differential rotation.  The relative importance
of the $\Omega$ effect over the $\alpha$
effect depends on the ratio of $C_\Omega$ and a similar parameter
$C_\alpha=\alpha/\eta_t k$ that characterizes the strength of the $\alpha$ effect.
Both $C_\Omega$ and the ratio $C_\Omega/C_\alpha$ would be underestimated in
an LES in which $\eta_t(k) k$ and $\alpha(k)$ are too big, so this suggests that
one would need to compensate for this shortcoming by increasing $\Omega$ to recover
cyclic dynamo action.  

Though this reasoning is generally robust, it is based on kinematic 
mean-field theory so its application to MHD LES must be made with care.
For example, MHD/LES convection simulations by \cite{BMBBT11} 
and \cite{Nelson13} demonstrated a transition from steady to cycling 
dynamos with both an increase in $\Omega$ and a decrease in the SGS component 
of the turbulent magnetic 
diffusivity, $\eta_t$.  This appears to be consistent with the 
mean-field arguments in the preceding paragraph.  
However, when the SGS diffusion was decreased in the
latter case (Nelson et al.\ 2013), the kinetic energy of the convection increased and the differential
rotation weakened due to Lorentz force feedbacks, implying an effective 
{\it decrease} in $C_\Omega$ for the cyclic case.  Furthermore, though the total 
magnetic energy was greater in the cyclic, low-dissipation case, the magnetic 
topology was more complex, with less energy in the mean (axisymmetric) fields.
This implies a relatively inefficient $\alpha$-effect.  Further analysis 
confirmed that despite the relatively weak $\Delta \Omega$ in the cyclic simulation, 
the primary source of toroidal flux was indeed the $\Omega$-effect, implying 
$C_\Omega / C_\alpha > 1$.

%AB.

%%%% Section 3 rewritten by WHM
\section{Small-Scale Dynamics: Dissipation, Reconnection and Kinetic Effects}
\label{sec:small}

LES methods as applied to HD have as a central goal
the ability to compute the dynamics of the resolved scales more accurately
under conditions in which the Reynolds numbers are too high for a fully resolved 
DNS computation.  Additional goals may be to compute the direct influence 
of the unresolved scales on the resolved scales (backscatter), or,
in the so-called energy equation approach, 
to track the transport of unresolved turbulence.
Specific formulations of LES 
relevant to the MHD case will be discussed in more 
detail in the following section. 
The complications that MHD introduces 
into small scale physics
become even more challenging
when MHD models  
are employed to approximate the 
dynamics of a low collisionality (or ``kinetic'') plasma.
These issues carry over into additional challenges for 
LES/SGS modeling.

In MHD the dissipation function 
(for simple resistivity and viscosity) is known, 
as in HD.
However the phenomenology of MHD dissipation is more complex
given that both current density structures and 
vorticity structures are available as sites of 
enhanced heating. Simulations have shown that this leads to a dependence 
of the ratio of kinetic to magnetic energy dissipation on the magnetic Prandtl 
number \citep{Bra09,Bra11,Bra14}, which is generally not reproduced by LES.

Not only is there an additional channel
for dissipation, but the nonlinear transfer 
of energy between velocity and magnetic field is known to 
be more nonlocal in scale than the transfer of total 
energy across scale \citep{Verma04,AlexakisEA05}.
The possibility that these effects come into play 
in varying proportions for 
MHD flows in differing parameter regimes is 
not only possible but likely, given for example
the well known differences in small-scale dynamics 
in the kinematic dynamo regime,  the Alfvenic turbulence
regime, and the large scale reconnection regime.
These differences reflect the degree of nonuniversality 
inherent in MHD behavior,
which has been demonstrated in a number of recent studies
\citep{LeeEA10,WanEA12}

Variability in the nature of the cascade
represents a challenge in 
developing LES for MHD 
since the correct modeling of important
sub-grid scale physics may be situation-dependent. 
However the challenge is even deeper
in the context of low collisionality astrophysical 
plasmas, even if MHD represents 
an accurate model at the larger scales. 
At the smaller scales, comparable to ion gyroscales or inertial 
scales, one expects MHD to break down
and give way to a more complete dynamical 
description, which is commonly referred to 
as the kinetic plasma regime.  
Processes occurring at the kinetic scales may resemble
analogous MHD processes, but may also differ
significantly in their detail. 
The LES developer in such cases may need to 
understand carefully whether the relevant processes to be incorporated
into sub grid scale modeling remain MHD-like, or 
if they are possibly influenced strongly by kinetic physics.

Two examples of processes that are potentially 
influenced by kinetic physics are dissipation of fluctuation energy 
and magnetic reconnection. These 
processes may lead for example, 
to electron/ion heating and nonthermal particle acceleration. 
In many space and astrophysical applications
of MHD, from the solar wind to black hole accretion disks,
these mechanisms can play a crucial role
for the global dynamics of the system, coupling microscopic 
and macroscopic scales.

One may also ask how the details of small
scale processes 
might have influence on the large-scale dynamics
that is the emphasis of LES. 
As long as the focus remains in the MHD range of scales,
in the usual way one may anticipate that 
energy transfer across scales will be almost independent of scale 
at high Reynolds number.
If an accurate estimation of the energy flux is available, 
it enables closure of the SGS problem before the dissipative scales 
are even encountered. 
This is a key step in the \cite{KarmanHowarth38}
similarity decay hypothesis, and is a familiar component 
of most HD LES. 
Even the MHD models can become more elaborate,
for example when 
there is a need to include backscatter effects
(as discussed above). Furthermore, 
for situations
that permit inverse cascade\footnote{Here we distinguish 
backscatter from inverse cascade, the latter being 
back transfer, or upscale transfer, driven by an additional 
ideal conservation law.}
this additional complexity in modeling energy transfer becomes
mandatory.
However when 
MHD models are employed for long wavelength description of 
kinetic plasma behavior, it transpires that
there are additional motivations for study of small scale effects
in building LES models. 
These potentially include:
\begin{enumerate}[(i)]
\item the requirement of following 
magnetic topology and connectivity, 
which may be influenced by small-scale processes such as magnetic
reconnection, as well as diffusive effects such as Field Line
Random Walk (FLRW); 
\item the requirement of computing 
test particle scattering and/or acceleration, in order 
to employ the models for study of
suprathermal particles, heat conduction or 
energetic particles such as cosmic rays and solar energetic particles;
\item the requirement of representing dissipation, heating
and more complex kinetic responses
(including in some cases radiative cooling), 
which may be regulated by the LES
fields.
\end{enumerate}
In each of the above problems the large scale MHD fields
and the cascade that they produce establish conditions
at the kinetic microscales, and the physically significant 
process -- reconnection, heating, particle acceleration, etc.,
follows as a response. 
It seems clear that an LES model that would include  
these effects must be more elaborate than one that focuses
mainly on energy flux. 

Pursuing a better understanding of the small-scale dynamics in MHD turbulence
in  the inertial range, and even smaller scale kinetic plasma dynamics in a turbulent medium,
has become a very active area of research in recent years.
This effort has been boosted by availability of high resolution 3D MHD codes
and kinetic plasma codes (fully kinetic, hybrid, and gyrokinetics),
as well as a wealth of new observational data regarding solar wind fluctuations
down to the electron gyroradius scale
\citep{AlexandrovaEA09,SahraouiEA09}. 
These studies have improved our theoretical understanding 
of the nature of the turbulence cascade and its effects
as it progresses from magnetofluid scales, to proton and electron kinetic scales.
The continuation of these advances is expected in the next few years
to provide a much improved basis for development of
SGS models that will enable a new generation of MHD and plasma
LES models. 
In the following, we shall
attempt to provide a brief overview of the current state-of-the-art as well 
as a discussion of key open questions regarding small scale dynamics.

\subsection{Do the Small Scales Matter?}

Before taking a detailed look at the small-scale dynamics that must be present in any turbulent MHD flow, we must first address a pressing question; do any of these details matter?  Recall the central premise of LES introduced in \S\ref{sec:intro}; Since the large scales generally dominate the turbulent transport and energy budget, these are the scales we are most interested in; why should we care about the small scales at all?

There are two answers to this question.  First, the small scale dynamics may influence the large-scale dynamics, often in ways we do not yet understand.  A notable example is global MHD simulations of magnetic cycles in convective dynamos.  Though remarkable progress has been made in recent years, such simulations are still quite sensitive to the nature of the SGS dissipation and the spatial resolution \cite{charb15}.  This is perhaps not a surprise, since the large-scale fields are intimately linked to the small-scale fields by, among other things, the topological constraints associated with magnetic helicity.  Large-scale dynamos rely on these linkages to generate magnetic energy and may thus be particularly sensitive to SGS processes.

More generally, the magnetic connectivity has the distinction of depending on microscopic properties 
such as reconnection activity, while clearly also having an influence on the large scale features 
of the problem at hand.  We now turn to the solar wind as another example that demonstrates this. 
During solar minimum conditions the fast solar wind is believed to 
emanate from polar coronal holes 
 while slow wind emerges from nearby regions outside the 
coronal holes, and perhaps from reconnection activity in coronal 
streamers. Stated this way it is possible that the 
boundary between fast and slow wind would be sharp, but this
is not observed; instead the transition is more gradual \citep{RappazzoEA12}. 
It has been suggested that this boundary is thickened
by random component interchange reconnection
\citep{LazarianEA12b,RappazzoEA12}
that causes there to be a band of field lines near the 
boundary that have a finite probability of connecting across this 
boundary due to dynamical activity.   
While high resolution
codes can simulate small regions near the boundary
to demonstrate this phenomenon, in an LES scenario 
it is doubtful that resolved scales would contain sufficient information 
to characterize this process. 
The resolved field lines would be nominal field lines, 
and if laminar, might maintain a sharp boundary at the coronal 
hole edges.
It would be a challenge for a refined LES/SGS model to
incorporate sufficient information about the space-time
structure of the unresolved fluctuations so that a 
model could be developed to represent 
both spatial randomization, due to 
field line random walk, and temporal randomization, due to 
potentially numerous unresolved reconnection sites.

It is not difficult to find other astrophysical plasma 
problems that depend on small scale, or even 
kinetic scale processes, while also having a significant impact on 
large-scale features.
Examples include small and large-scale dynamos (\S\ref{sec:dynamo_theory})
as well as the relative level of electron, proton 
and minor ion heating
in the solar wind or in black hole accretion disks. 
Here, the small-scale physics plays a critical
role in determining the overall magnetic topology, radiative signatures,
and thermodynamics of the system, with significant large-scale 
observable consequences.

The second answer to the question of ``why should we care about the small sales at all?'' is that the small-scale dynamics can potentially have observable consequences that are regulated by the large-scale flows and fields.  A notable example is particle acceleration in solar flares and interplanetary shocks.  Sharp gradients in large-scale fields promote small-scale reconnection that often produces a non-thermal spectrum of high-energy particles.  These solar energetic particles (SEPs) are an important component of space weather, with potential socio-economic consequences.  The small-scale reconnection that produces SEPs also dissipates energy (and other global quantities) and reshapes the magnetic topology.  Thus, it may be necessary in some situations to take particle acceleration into account when devising high-fidelity SGS models.  In such cases an energy equation formalism would be desirable in order to compute particle diffusion coefficients.

Yet, there are many HD and MHD applications when a simple dissipative SGS model will suffice.  Here the large-scale dynamics is insensitive to the small-scale dynamics, provided that the Reynolds and magnetic Reynolds numbers are high enough to resolve coherent structures and capture self-similar cascades.  A notable example here is solar granulation (see \S\ref{sec:solar_granulation}).  In this case, one would be satisfied with relatively simple LES models, such as ILES (see \S\ref{sec:lesmhd}). 

In order to assess whether or not a sophisticated SGS model is needed, and in order to devise such a model when necessary, one must have a comprehensive understanding of the fundamental physical processes that operate at small scales, and how they influence large-scale dynamics.  This is where we now turn.

\subsection{Physics of the small-scale cascade}\label{sec:ssc}

Laboratory plasmas provided the 
first quantitative indication that 
MHD turbulence is anisotropic relative to the large scale 
magnetic field direction 
\citep{RobinsonRusbridge71,ZwebenEA79}, 
generating spectral or correlation
anisotropy with stronger gradients 
transverse to the magnetic field 
and weaker parallel gradients.
Simulations in both 2D and 3D 
demonstrated the dynamical basis for this effect:
propagation of fluctuations along the magnetic field interferes
with parallel spectral transfer, while perpendicular transfer
remains unaffected 
\citep{ShebalinEA83,OughtonEA94}.
Correlation anisotropy of the same type was found
to operate relative to the {\it local} magnetic field
\citep{ChoVishniac00,MilanoEA01}.

Spectral anisotropy generates a distribution
of excitation in wave vector such that 
average perpendicular wavenumbers 
are greater than average parallel wavevectors, i.e., 
$\bar{k}_\perp > \bar{k}_\parallel$,
relative to the global field.
The degree of anisotropy becomes greater at smaller 
scales, so, for example the anisotropy of 
$\nabla \cross {\bf B}$ exceeds that of $\bf B$
\citep{ShebalinEA83}.
Moreover, local correlation anisotropy measured by conditional 
structure functions
\citep{ChoVishniac00,MilanoEA01}
is greater than 
global anisotropy.

Another familiar type 
of anisotropy that emerges 
in plasma turbulence at MHD scales 
is polarization (or variance) anisotropy.
In this case one finds that mean square value 
of each component of 
the fluctuations perpendicular to the mean magnetic field
is larger than the mean square parallel component.
This condition emerges naturally in Reduced 
MHD treatments
of tokamak plasma devices, in which the aspect ratio of the 
device plays a key role 
\citep{KadomtsevPogutse74,Strauss76}
%%(Kadomtsev \& Pogutse, 1975;Strauss, 1976) 
and the resulting nonlinear
dynamics is both transverse and incompressible, and 
also requires spectral anisotropy with $k_\perp \gg k_\parallel$
as discussed above.  
Later it was shown that Reduced MHD (RMHD) and its transverse 
fluctuations may be derived by elimination of fast magnetosonic
and Alfvenic timescales in solutions of the 
full 3D compressible MHD equations with a strong mean magnetic field
\citep{Mont82-strauss,ZankMatt92a}.

It is noteworthy that the properties of 
low frequency, high-$k_\perp$, incompressible fluctuations
with transverse polarization, equates in wave vocabulary to 
dominance of the oblique Alfv\'en mode, and suppression of the 
magnetoacoustic modes.
This characterization of fully developed incompressible
inertial range  MHD turbulence -- 
consisting primarily of a highly oblique
spectrum of  transverse 
fluctuations has provided a basis for models
of plasma turbulence by a number  of authors
\citep{MontTurner81,Higdon84,Goldreich}.

While there are 
a number of differences in these 
formulations, they have in common
that MHD turbulence gets more and more anisotropic at smaller scales. 
One approach \citep{Goldreich}
introduced the term  ``critical balance,'' 
to describe the fate of weakly interacting Alfv\'en waves 
that produce 
perpendicular spectral transfer until 
nonlinear (perpendicular)
eddy and linear (parallel) Alfv\'enic timescales 
become equal. This establishes 
a relationship
between perpendicular and parallel wave numbers that 
is characterized by $k_\parallel\propto k_\perp^{2/3}$.
As a consequence, one finds $k_\parallel\ll k_\perp$ at small scales.
The same relationship is found in the earlier turbulence theory  
\citep{Higdon84}
based on quasi-two dimensional or RMHD spectral transfer 
\citep{Mont82-strauss,ShebalinEA83} 
except that the RMHD turbulence energy is mainly confined to the 
region of wave vector space in which the 
nonlinear time scale is less than 
the linear wave timescales. The 
relationship $k_\parallel\propto k_\perp^{2/3}$,
is common to both, if the wavenumbers are regarded as averages
of the energy spectrum in the inertial range.
In any case the preference for perpendicular spectral transfer 
\citep{ShebalinEA83}
 appears to be a robust result in MHD 
turbulence and should be considered in SGS modeling
when there is a uniform magnetic field or a 
very large scale magnetic field present. 

In addition to the energy spectrum, the inertial 
range in MHD turbulence is characterized 
by additional correlations.
The velocity and magnetic fields are typically correlated in 
direction with the sense of correlation
coherent within patch-like regions of real space
\citep{MilanoEA01,MatthaeusEA08}. 
%% L. J. Milano, W. H. Matthaeus, P. Dmitruk \& D. C. Montgomery, Phtys. Plasmas 8, 2673 (2001)
%% W. Matthaeus, A. Pouquet, P. D. Mininni, P. Dmitruk \& B. Breech, Phys. Rev. Lett., 100, 085003 (2008)
A complementary idea is that 
the alignment increases systematically with decreasing scale
\citep{Boldyrev06,MasonEA06}. 
%% Boldyrev, PRL, 96, 115002 (2006); Mason et al, PRL 97, 255002 (2006)
It is also documented that turbulence produces patchy, 
localized correlation of other kinds in MHD 
\citep{ServidioEA08},
and at least some of these appear to be related to the tendency for turbulent 
relaxation 
\citep{TingEA86,StriblingMatthaeus91}
to proceed locally in cellular regions, such as flux tubes, as a faster, 
intermediate step towards global decay and relaxation.
The types of correlations produced locally and rapidly in this way include 
(but are not limited to), not only the Alfv\'enic correlation (velocity and magnetic
field),  but also the Beltrami correlation (velocity and vorticity) and the 
force free correlation (magnetic field and electric current density). 
All lead to depression of nonlinearity in the inertial range of scales, 
as seen in the emergence of Beltrami correlation in HD
\citep{PelzEA85}.
%%(Pelz et al, 1985) 
%%  

It is not entirely clear how or whether these additional correlations
should be included in LES/SGS modeling of the smaller scale MHD cascade. 
On the one hand, the diversity in possible long-term relaxed states 
suggests dominance of different relaxation processes for different 
parameter regimes. For example, to achieve global dynamic 
alignment, any excess mechanical or magnetic energy would 
need to be dissipated. Similarly, in order 
to achieve global selective decay of energy with 
constant helicity 
\citep{MontgomeryEA78,MatthaeusMontgomery80},
also known as Taylor relaxation \citep{Taylor74}, 
would require that 
mechanical energy be entirely dissipated while magnetic energy remains. 
Presumably these alternative decay prescriptions place different 
requirements on the nature of dissipation models.
Dynamo action with injected mechanical helicity at intermediate 
scales also places requirements on transfer and dissipation rates of 
energy, magnetic helicity and kinetic helicity 
\citep[e.g.][]{Brandenburg01,BN11,BS05}.
%% Brtandeburg, ApJ, 550, 824 (2001)
%% Astrophysical magnetic fields and nonlinear dynamo theory, Axel Brandenburg, Kandaswamy Subramanian, Physics Reports 417 (2005) 1–209
On the other hand if the 
processes being modeled are principally dependent 
on the decay rate of energy, it may be possible to 
define energy fluxes 
with relatively simpler  prescriptions, such as by 
partitioning transfer between 
direct and inverse cascade rates.  
How these issues will influence improved and accurate LES/SGS
models for MHD in the future is a current research-level problem 
that is intimately tied in with prospects for universality in MHD 
turbulence, or perhaps universality within classes of MHD behavior.

While it will likely be necessary to learn more about MHD and kinetic scale 
cascades to build more complete models, it is noteworthy 
that considerable theoretical progress has been made, 
including computations,
by assembling turbulence models that 
may lie somewhat outside of a strictly-defined LES
concept. These models typically have concentrated on 
selected effects that are thought to be dominant 
for the chosen problem. 
Examples of such models are 
mean field electrodynamics 
\citep{Moffatt,KrauseRaedler}
often used in dynamo theory, 
Reynolds averaged MHD models such as those used for 
solar wind modeling 
\citep{UsmanovEA14}
and hybrid models based
on multiple scale analysis and Reynolds averaging
\citep{YokoiEA08}.
In any of these models, we should note that a turbulent resistivity would have essentially 
the same effect as an ``anomalous'' resistivity, by which we mean a contribution to resistivity 
due to small scale (and also unresolved) kinetic plasma effects. 
This is also an area that has been well studied \citep[e.g.][]{biskamp00}.

%%(Yoshizawa, 19XX, Yokoi \& Yoshizawa, 1992;
%% Yokoi et al, 2008).
%% Yokoi, Nobumitsu; Rubinstein, Robert; Yoshizawa, Akira; Hamba, Fujihiro J. Turbulence, 9, 37, 2008
%%  Journal of Turbulence Vol. 9, No. 37, 2008, 1–25
%% A turbulence model for magnetohydrodynamic plasmas
%% Nobumitsu Yokoi, Robert Rubinstein, Akira Yoshizawa, Fujihiro Hamba
%%
As an example of a non-traditional LES approach, \cite{YokoiEA13} describe a novel self-consistent mean-field theoretical model of turbulent MHD reconnection, highlighting cross-helicity dynamo effects.  In this, essentially sub-grid, model the effects of small-scale turbulence are represented by two additional terms in the Ohm's law: one proportional to the turbulent energy density and describing standard effective turbulent resistivity, and the other, new term, proportional to turbulent cross-helicity $W = \langle{\bf u'} \cdot {\bf b'}\rangle$ and the large-scale  vorticity ${\bf \Omega} = \nabla \cross {\bf U}$.  Though this model appears to capture the influence of small-scale turbulence on large-scale reconnection \citep{YokoiEA13}, \cite{grete15} found that it does not perform well for supersonic MHD turbulence, where it fails to reproduce the turbulent electromotive force (EMF) obtained from high-resolution ILES (standard eddy diffusion models also fail in a similar way).  More work is needed to determine its viability in different circumstances.  Indeed, this applies to all SGS models; to the extent that it is feasible, their validity and scope should be evaluated by comparing them to high-resolution DNS/ILES \citep[e.g.][]{grete15,meheu15} and/or to kinetic plasma simulations. 

Having introduced some prominent features
of the physics of MHD turbulence at small scales, we will
now focus on some recent findings from respective studies
that are relevant to the fate of the MHD cascade at smaller scales. 
A central 
issue for many applications 
in space and astrophysics is how the cascaded energy 
is actually dissipated, and in some astrophysical 
systems, eventually radiated away.  
For the present purposes the 
notion of dissipation may be 
described as the irreversible 
conversion of large scale or fluid scale energy
into microscopic kinetic degrees of freedom.
Important questions that have been recently addressed 
in this area include the 
response of test particles to MHD electromagnetic fields, 
kinetic effects including dissipation of cascaded
MHD fluctuations and the response in the form of heating, 
and the role of magnetic reconnection,
current sheets and tearing, 
and the associated macroscopic effects of changes in magnetic 
topology and connectivity. 

\subsection{Energization and transport of test particles}

The most primitive model of kinetic response to 
MHD-scale fields is given by the test-particle
approximation in which the trajectory of 
individual plasma particles is assumed to be 
determined by the Newton-Lorentz force law, neglecting all 
feedback of the particle motion on the rest of the plasma 
or on the electromagnetic fields.
The basic physics of acceleration, 
scattering and transport, especially of 
suprathermal and 
energetic particle populations, 
is often discussed in a first approximation
using a test particle 
approach \citep[e.g.][]{Bell78,Jokipii66}. 
Not only are test particle studies useful in understanding 
energy dissipation, but in some cases, e.g., cosmic rays and solar energetic 
particles, it is the response of the
test particles to the large scale fields, and the subgrid scale fields,
that is the essential output of the research. 

A self-consistent model
extending beyond test particles  
is needed for accurate 
representation of the 
effects on dissipation
of slower populations of plasma particles, 
say, those moving at a few Alfv\'en speeds or less.
Nevertheless, in spite of its shortcomings, 
the test particle approach, implemented in concert with 
MHD computations, has been valuable for  
investigation of potential mechanisms of energization and 
dissipation prior to emergence of computational capacities 
that enable equivalent self-consistent 
kinetic modeling. 

A  good example of this is the
use of test particles in the elucidation of the role of reconnection and 
turbulence in energization of suprathermal particles. 
Spectral methods, having favorable resolution properties for
turbulence,  were able to describe the interplay of test particle energization
and nonlinear reconnection at a relatively early stage
\citep{AmbrosianoEA88}.
In the presence of strong fluctuations, reconnection does
not settle in to smooth solutions anticipated from tearing mode  theory,
and instead remains unsteady and bursty, and when the Reynolds 
number at the scale of the dominant current sheets
exceeds a few hundred, the fluctuations lead to multiple small 
magnetic flux structures, or secondary islands
\citep{MatthaeusLamkin85,MatthaeusLamkin86,Biskamp86}.
 This subject is revisited in more detail below in 
section \ref{sec:recon}.

Here we note simply that 
such structures can entrain or temporarily trap test particles, 
and are strongly associated with the most efficiently energized particles.   
This entrainment and energization 
was found to occur between magnetic X-points and O-points, 
as was later found in much greater detail and realism 
using high resolution kinetic plasma codes 
\citep[e.g.][]{DrakeEA06}. 

It is clear that 
even a simple model employing 
MHD simulation fields and test particles
can begin to identify kinetic effects 
beyond simple energization. 
Studies showed that small gyroradius particles 
(e.g., electrons) 
tend to be accelerated in the direction along the electric 
current sheets, that is, parallel 
acceleration, while heavier particles (protons, etc) 
are energized in their perpendicular velocities 
\citep{DmitrukEA04}.
Self-consistent kinetic simulations also were able to find this 
effect, and in fact it is now understood through plasma simulation
that the regions {\it in and near} current sheets
are sites of enhanced kinetic effects such as suprathermal particles, 
temperature anisotropies, large heat flux, and in general non-Gaussian features of the 
proton distribution function \citep[e.g.][]{ServidioEA12,KarimabadiEA13}.

More recent test particle studies that employ weakly 3D RMHD 
simulations 
\citep{DalenaEA14} 
suggest that energization of a single species of test particle 
progresses through at least two stages in the presence of a strong guide field
with nearly two dimensional low frequency 
fluctuations: 
First, at lower energies the particles are energized in their parallel velocities,
and mainly while entrained near reconnection sites inside of current sheets
and in essentially in accord with the classical neutral point acceleration mechanism.
This is also sometimes called ``direct acceleration.''
As suprathermal energy grows and the gyroradii become 
larger than the typical thickness of the current sheets, 
the energization of test particles continue, but with 
enhancement of perpendicular velocities.
This has been described as a ``betatron'' process
associated with an inhomogeneous 
perpendicular electric field found near to, but outside of strong reconnection sites
\citep{DalenaEA14} .  
This test particle result 
provides more detail on 
earlier results on acceleration in turbulence
\citep{DmitrukEA04,Chandran10}.
   
The marriage of test particle 
studies with high resolution MHD simulation 
has led to a number of insights and questions
that are of relevance to ongoing efforts to develop SGS models 
for MHD and plasmas.
For example: 

$\bullet$ When are the 
processes of cascade, reconnection, 
test particle energization, and dissipation,
related? 
While it is fairly clear that a broad-band cascade 
requires reconnection to occur at various scales
along the way, the fact 
that test particles respond to the 
associated inhomogeneities suggests
that dissipative processes 
may be intertwined  in this process. 

$\bullet$
The topology of the magnetic field
becomes fuzzy when there are numerous 
small secondary islands, so trapping, reconnection,
coalescence and  particle 
energization
will in general not be explicitly resolved in an SGS/LES 
scheme for a large system. 
Most likely these
features will need to be understood well 
enough to develop a statistical or phenomenological
model.

$\bullet$ 
If tracking test particle populations
at a statistical transport level
remains a scientific priority in an LES context, 
then a requirement will be to follow parameters needed for 
SGS transport models, such as SGS energy,
characteristic length scales, and possibly spectral features 
such as anisotropies, e.g., to capture possible resonances. 

\subsection{Kinetic effects, dissipation processes and heating}\label{sec:kinetic}

Once cascading
fluctuations reach scales
at which kinetic effects become 
important, MHD is no longer applicable.
In practical terms, this means that for 
 an ion-electron plasma, when the cascade arrives
at scales as small 
as either the ion gyroradius scale, $\rho_i$, 
or the  ion inertial scale $d_i = V_A/\Omega_{cp}$,
kinetic effects become  important and even dominant. 
For wavenumber $k$ the corresponding kinetic 
range is indicated by $k\rho_i \geq 1$ 
or $kd_i \geq  1$. 
To retain effects like finite Larmor radii and Landau damping in this regime, one has to
employ a kinetic description\footnote{By kinetic description 
we mean a dynamical description of the 
plasma that involves only the one-particle distribution function
which depends on velocity, position, and time.}.
Since in space physics and astrophysics 
we are often dealing with low density plasmas 
for which the collisionality is very weak, it is important to keep in mind that 
some type of effective 
``collisions'' will inevitably 
cause departures from 
an idealized model such as the Vlasov-Maxwell equations.
Fundamental effects  such as an increase of 
system entropy and relaxation towards  
thermal equilibrium, will rely on the presence of these formally 
small 
contributions to the kinetic equations
\citep{Klimontovitch97,SchekochihinEA09}.
%%(Klimontovich, 1997; Schekochihin et al, 2009). 
In any case,
given the enormous
computational cost of nonlinear kinetic simulations in six phase-space dimensions, 
such studies often 
fall into the category of ``extreme computing.'' 
Results on turbulence energy dissipation and relaxation in turbulent 
plasmas, employing  
particle-in-cell (PIC) Vlasov code and Eulerian Vlasov 
codes, are just starting to appear in the literature 
\cite[e.g.][]{DaughtonEA11,ServidioEA12,KarimabadiEA13,HaynesEA14}.

In light of the impressive continued growth of supercomputing power as we head towards the exascale era, it may be expected that kinetic simulations will be at the forefront 
of research into the fate of cascaded energy at kinetic scales
in the years to come. Meanwhile, various 
reduced models 
are being used to complement fully kinetic studies.
These include hybrid 
(fluid electrons and kinetic ions), gyrokinetic 
and gyrofluid models, and an array of fluid models that 
contain some kinetic effects
(Hall MHD, multifluid, Finite Larmor radius MHD, etc.)

\begin{figure}[t!]\begin{center}
\leftline{\includegraphics[width=0.4\columnwidth]{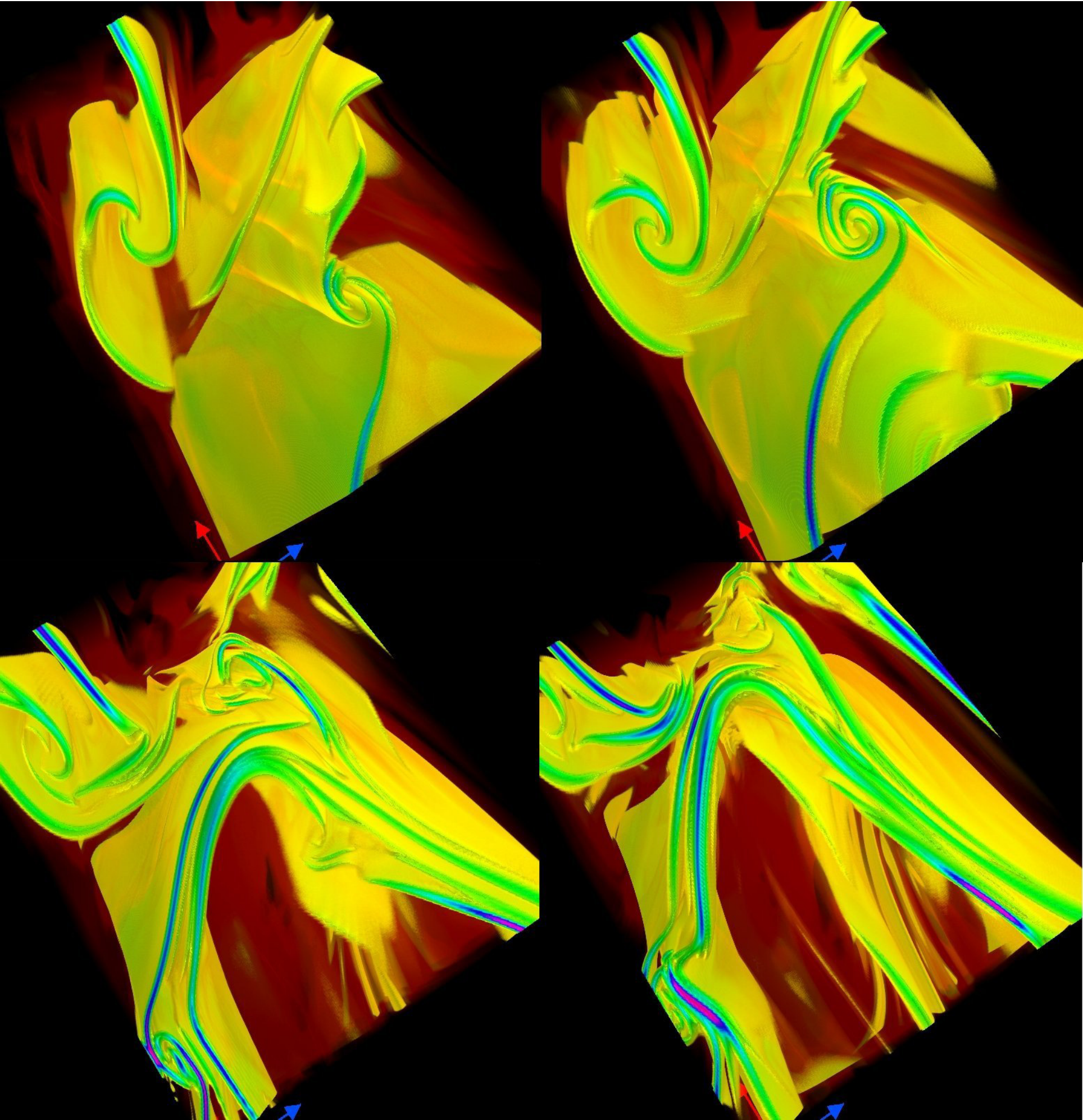}}
\vskip-0.4\columnwidth
\rightline{\includegraphics[width=0.5\columnwidth]{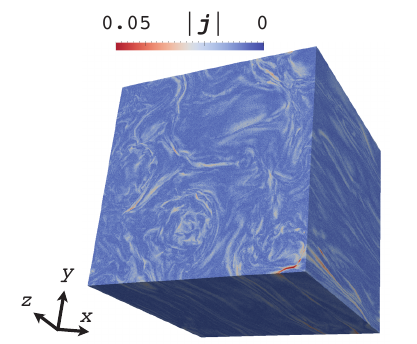}}
\end{center}
\caption{(Left) Volume rendering of magnitude of current density $J$
in the same small region of a high resolution 
3D MHD simulation at four different times, showing complex 
spatial structure and evolution in 
time \citep[adapted from][]{MininniEA08}. 
(Right) Volume rendering of $J$ from a $2048^3$ PIC simulation of plasma turbulence, 
in a periodic box of side 83.8 thermal proton gyroradii. Again, 
fine scale structure is evident, now at kinetic scales.
(Courtesy of V. Roytershteyn, to be published).}
\label{fig:3dcurrents}.
\end{figure}

Fully kinetic 3D PIC simulations of turbulence and reconnection 
have recently revealed interesting details about kinetic response 
to cascading MHD scale fluctuations.  
In 3D, ion-scale current sheets spontaneously develop turbulence through various 
instabilities, producing a chaotic 3D magnetic field structure \citep{DaughtonEA11}.
Examples of fine scale current structures in high-resolution MHD and 
kinetic simulations are shown in Fig.\ \ref{fig:3dcurrents}.
Interestingly, however, both the reconnection rate and the mechanism for 
breaking the frozen-flux law seem to be unaffected, being close to those 
obtained in 2D simulations. 
Numerical experiments with different types of initialization, 
such as velocity-shear-driven kinetic turbulence
\citep{KarimabadiEA13} 
show that 
current sheets and intermittency 
can form 
in both 2.5D PIC simulations as well as 3D PIC simulations.
Furthermore, kinetic activity of various types, including heating,
appears to be concentrated near sheet-like current structures
\citep[see also][]{ServidioEA12,WanEA12a}, 
which lends credence to the emerging idea that a large fraction of 
kinetic heating may occur in or near current sheets and related structures. 
(For the MHD analogue of intermittency associated 
with current sheets, see section 3.4 below.) 
Some wave activity is also identifiable 
in the above examples, although in terms of the partitioning of 
energy, this seems to be the exception rather the 
rule when broad-band turbulence develops 
at kinetic scales \citep{ParasharEA10,VerscharenEA12,KarimabadiEA13}.
Kinetic scale complexity and intermittency is also observed
using recent high resolution observations in the solar wind
\citep{PerriEA12,WuEA13}. 

Even with recognition of this complexity at kinetic scales, 
efforts continue to analyze the type of wave mode, or perhaps several types,
that might be viewed in some sense as the elementary excitations from 
which the turbulence is constructed.  There are actually
several approaches 
that may be described as a wave turbulence approach.
On the one hand there is the formal 
weak turbulence theory \citep[e.g.][]{GaltierEA00}
that considers the cases in which the leading order   
dynamics is that of propagating waves that obey, to a first
approximation, a dispersion relation that assigns a frequency to each 
wavevector. Another view is that the distinguishing
characteristics of wave 
modes are the polarizations and wave vector directions, 
which suffice to establish an identification.

Substantial research has been devoted to models
 that are built upon the premise of wave modes that couple to 
produce wave turbulence in the nonlinear regime. 
One class of such models, defined in the MHD regime, is 
the Goldreich-Sridhar theory \citep{Goldreich},
(also called ``critical balance'' after one particular assumption that 
is made in the theory; see Sec.\ \ref{sec:ssc}). This theory assumes that 
all possible excitations are Alfv\'en modes, having polarizations
strictly perpendicular to the applied mean magnetic field. The usual argument 
is that other wave modes evolve semi-independently so that the 
evolution of the Alfv\'en waves and their mode-mode 
couplings can be computed independently of the 
magnetosonic wave turbulence. This idea is also routinely carried
over to kinetic regimes,  in which it is assumed that 
distinct wave modes, such as kinetic Alfv\'en waves (KAW),
or whistlers, will evolve independently.  Numerical 
evidence is usually invoked to support this assumption,
but the idea remains somewhat controversial.
For example, 2.5D kinetic simulations that are initiated
with Alfv\'en modes, i.e., zero parallel variance, appear to 
generate parallel variance fairly rapidly, although 
at a lower level. Thus, in wave terminology,
magnetosonic mode turbulence is generated by 
Alfv\'en mode turbulence within the time span of the
current generation of simulations which are relative short due to finite 
computational resources, typically less than, say, 1000 proton
cyclotron periods. 
Furthermore it is well known that the parallel variance 
component of solar wind turbulence is small but nonzero, 
as in the famous ``5:4:1'' observations by 
\cite{BelcherDavis71}
using Mariner data.

Goldreich-Sridhar turbulence, which is purely
Alfv\'en mode, 
evolves from  
a wave state through standard weak turbulence couplings
towards the critical balance state, provided that the 
zero frequency modes (purely 2D nonpropagating fluctuations) 
are absent or very nearly absent.
This results in a wave turbulence that is highly oblique,  
with mainly near-perpendicular wave vectors involved
in the dynamics.
 At small scales approaching the kinetic range, 
the oblique Alfv\'en modes in Goldreich-Sridhar
theory naturally 
go over to Kinetic Alfv\'en waves 
\citep{Hollweg99},
which have received substantial 
attention recently in solar wind observations
%%(Bale et al, 2005; Sahraoui et al, 2008; Salem et al, 201X).
\citep{BaleEA05,SahraouiEA10}

Another wave mode discussed in connection with
wave turbulence and a possible role in the
kinetic range of solar wind dynamics is the whistler mode
\citep{HughesEA14},
which generally is at higher frequencies than the KAWs and 
probably has lower amplitude in the solar wind, but may still play a role in 
the operative dissipation mechanisms. 

Much of the debate concerning relative roles of wave 
modes has taken place in the 
context of recent high resolution 
measurements of fluctuations in the dissipation range of solar wind turbulence 
\citep{BaleEA05,AlexandrovaEA09,SahraouiEA09,SahraouiEA10}.
First, it was observed that the electric and magnetic
field fluctuations as well as the density fluctuations in the scale range 
of $\rho_i^{-1}\ll k_\perp\ll\rho_e^{-1}$ 
display power law (not exponential) spectra. 
While the knee at $k_\perp\rho_i\sim 1$ was originally 
attributed to some form of damping, e.g., 
proton cyclotron damping or Landau damping of 
kinetic Alfv\'en waves (KAWs), it was later suggested 
that the observed power law exponents can be explained solely on 
the basis of dispersion effects at these scales 
\citep{StawickiEA01}.  Consequently, this scale range is now sometimes 
also called a ``dispersion range''.
Neglecting cyclotron absorption at the proton resonance, 
it may be that significant ion/electron 
dissipation sets in, respectively,
at sub-ion scales, $k_\perp\rho_i\gg 1$, and electron scales, $k_\perp\rho_e\sim 1$. 
Beyond the electron scales one might expect the
occurrence of exponential spectra 
\citep{AlexandrovaEA09}
although 
there are also observations consistent with yet another power law 
\citep{SahraouiEA09}. 
Clearly, the physics of this entire sub-ion scale range is of interest
to understanding the heating of turbulent space and astrophysical plasmas. 
The relative importance of ion and proton kinetic mechanisms 
that might give rise to dissipation is likely determined by turbulence
amplitude \citep{WuEA13b} 
in addition to kinetic plasma parameters
such as 
the ion-to-electron temperature ratio $\tau$
and the plasma beta $\beta$.

%Quantitative predictions for the scaling exponents and relative
%heating rates in the dissipation range have recently been obtained
%within the framework of nonlinear gyrokinetics. The power law
%exponents for the electric and magnetic field fluctuations exhibit a
%remarkable level of agreement with the solar wind data from spacecraft
%observations.  Taking into account that the small-scale dynamics in
%gyrokinetics involves (only) KAWs, this is a clear indication that the
%solar wind data at sub-ion scales are compatible with KAW
%turbulence. Further support for this notion is provided by the fact
%that the measured density fluctuation spectra are in line with the
%physics of KAWs, but not whistler waves which have been proposed as
%another possible explanation.  More work will have to be done to
%confirm and refine this emerging scenario, in particular with respect
%to the role of non-Maxwellian velocity space features as well as
%mirror and firehose instabilities (via fully kinetic nonlinear
%simulations).

Standard approaches to studying the physics of the 
kinetic range of turbulence are Lagrangian PIC and Eulerian 
solutions of the Vlasov equation. 
as well as the hybrid (fluid electron) variants of each of these.
However, there have been special reduced models 
that have emerged that include interesting subsets of the relevant physics.
Nonlinear gyrokinetic theory 
\citep[see][and references therein]{SchekochihinEA09}
has been developed in the 
context of magnetic confinement fusion research since the early 1980s, 
and today it serves as the workhorse for computations in tokamak research.
An adaptation of gyrokinetics, 
as embodied e.g.\ in the GENE \citep{jenko00}
and AstroGK \citep{HowesEA08} codes,
includes a subset of possible
gyrokinetic effects, and  
has been proposed as a model for turbulence investigations in weakly collisional, 
strongly magnetized space and astrophysical plasmas from the inertial range through the 
ion and electron kinetic ranges. 
%As currently implemented,  these
%formulations are  
%perturbative about an isothermal Maxwellian, 
%with the same small parameter 
%associated with the ordering of 
%$\delta b/B_0$, $k_\parallel/k_\perp$ and the 
%dynamical time scale normalized to the 
%proton gyroperiod. Therefore 
%high frequency waves are also ordered out of
%the description. 
The main limitation of standard gyrokinetic theory is that
it is based on a low-frequency 
(compared to the particles' cyclotron motion) ordering, 
assuming a decoupling the fast gyrophase dependence 
from the slow gyrocenter dynamics. 
Notably, this version of gyrokinetics
lacks cyclotron resonance and therefore maintains
particle magnetic moments, thus 
placing it at odds
with some theories of solar wind and coronal heating.
We note in passing that this constraint can be removed if
necessary, leading to extended versions
of gyrokinetics \citep{qin00}.  In any case,
the formulation does include the physics of 
kinetic Alfv\'en waves, and goes over to 
Reduced MHD in appropriate limits. 
As such gyrokinetics 
is well suited to describe the 
Goldreich-Sridhar cascade. 
Gyrokinetics provides a computationally efficient 
method to study certain problems, 
and it has been argued that it does capture the physics 
needed to describe the observed turbulence
\citep{HowesEA08}
 and heating \citep{TenBargeEA13};
 this however remains a topic of lively discussion.
 It is also worth noting that gyrokinetics itself has been treated using
 an LES approach \citep{morel11,morel12}.

Gyrofluid theory is an attempt to reduce gyrokinetics to a multi-fluid approach
via calculating moments and closing the resulting hierarchy of equations by providing suitable closure
schemes \citep{hammett90,PassotEA12}. 
Kinetic effects like finite Larmor radii and linear Landau damping can be retained with
reasonable accuracy, provided that the closures are carefully constructed. 
Also pioneered in the context
of magnetic confinement fusion research in the 1990s, 
gyrofluid models have more recently 
been tailored and applied
to various space and astrophysical problems.
The 
development and refinement of this approach is a subject of on-going research.

It is at present 
unclear which model or models will provide 
what is needed for development of effective LES for low collisonality plasmas. 
If we knew which processes were important,
then selection of the appropriate 
reduced description models such as 2.5D kinetic codes, 
hybrid codes with fluid electrons, gyrokinetic codes or gyrofluid codes,
may provide the efficiency needed to arrive at the needed answers 
more quickly. However if those processes need to be identified, 
then more demanding 3D fully kinetic Vlasov or PIC representations
may be required. 

\begin{figure}[t!]\begin{center}
\includegraphics[width=\textwidth]{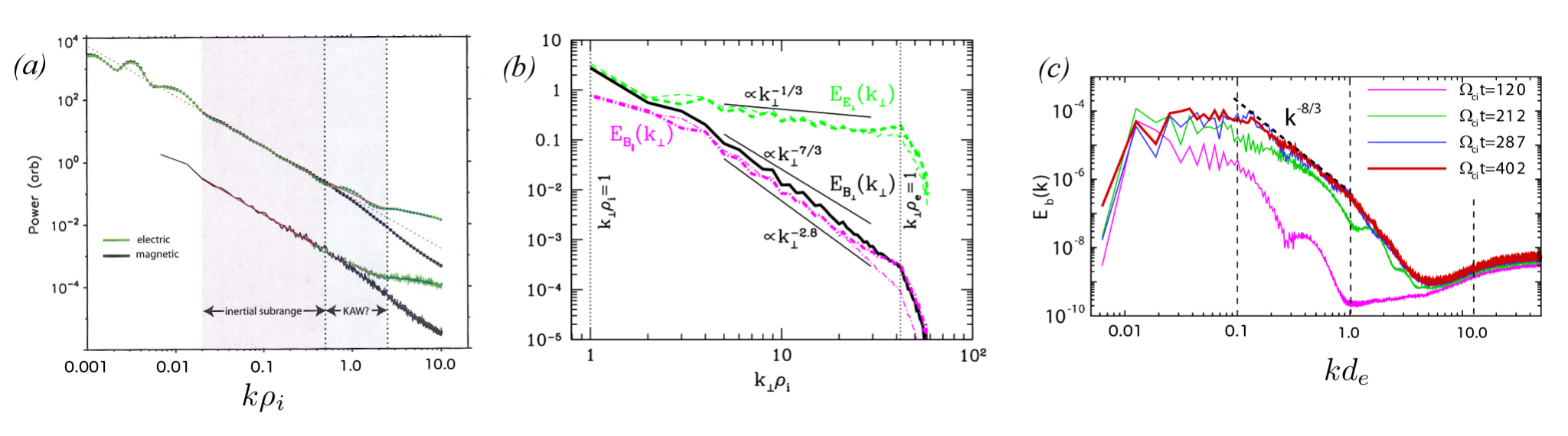}
\end{center}
\caption{($a$) Magnetic and electric field spectra in the 
solar wind obtained from in situ measurements by the Cluster spacecraft,
from \citet{BaleEA05}.  Corresponding spectra from ($b$) gyrokinetic 
simulations \citep{HowesEA08}, and from ($c$) electromagnetic PIC 
simulations \citep{KarimabadiEA13} are also shown for comparison.
In frames ($a$) and ($b$), wavenumbers are normalized by the thermal 
ion gyroradius $\rho_i$; in frame ($c$) the wavenumbers are normalized
by the electron inertial scale $d_e = c/\omega_{pe} = V_A/\Omega_{ce}$.
\label{fig:swspectra}}
\end{figure}

 We close the section with a few remarks concerning the 
prediction of turbulence spectra in the kinetic range and the associated
ambiguities of identifying dissipation mechanisms.
Within a wave turbulence framework, one may establish a contrast 
between expectations of KAW-mode turbulence and 
whistler-mode turbulence. 
Solar wind observations indicate a dispersive effect near the ion 
inertial scale that has been associated with KAWs 
\citep{BaleEA05} and with gyrokinetics \citep{HowesEA08}; 
see Fig.\ \ref{fig:swspectra}.
However this can qualitatively be explained by any 
dispersive processes that include a Hall effect 
\citep{MatthaeusEA08b}, 
so this is not a conclusive observation. 
On the other hand, the KAW and whistler dispersion relations, 
examined in detail in the observations, seem to favor KAWs more 
than whistlers in kinetic wave number ranges above $k \rho_i \sim 1$
\citep{SahraouiEA09}.
However a strong conclusion cannot be derived from this concerning dissipation 
processes because, first, we do not know that the dissipation occurs at 
these wavenumbers -- if it were to occur at much higher electron wavenumbers, 
there may be whistlers or other higher frequency waves that actually 
do the dissipating; and second, we do not know for certain 
that a wave turbulence treatment is even appropriate. 
Figure \ref{fig:swspectra} compares solar wind observations
with both gyrokinetic and kinetic simulations.  In all cases the
magnetic spectra break to something steeper than the inertial 
($-5/3$) range at or near ion kinetic scales. The latter may be 
the thermal ion gyroradius $\rho_i$ or, especially at low plasma beta, 
the ion inertial scale $d_i = c/\omega_{pi} = V_A/\Omega_{ci}$. (Note that, when the 
plasma $\beta$ is unity, $\rho_i = d_i$.) The PIC result shows a similar 
spectrum (possibly closer in slope to $-8/3$ than to $-7/3$)
with wavenumber normalized by electron inertial 
scale $d_e = c/\omega_{pe}$. Note that 
$d_e/d_i = \sqrt{m_e/m_i}$. Evidently the 
spectra themselves do not strongly differentiate between models. 
Additional examples from computations are shown in \citet{MatthaeusEA08},
\citet{SahraouiEA09}, and \citet{AlexandrovaEA09}.

If stochastic acceleration and scattering of orbits 
\citep{DmitrukEA04,Chandran10,DalenaEA14} 
in or near current sheets absorbs substantial fluctuation 
energy, then one may have to look beyond linear damping of wave modes 
for an effective dissipation mechanism. Other heating 
processes are also evidently available near turbulent 
reconnection sites and other coherent structures  that are seen in very high resolution 
kinetic simulations \citep{KarimabadiEA13,WanEA12}.
Furthermore, other kinetic instabilities 
(such as firehose and mirror instabilities) 
and the fluctuations they produce
may be particularly active near coherent structures 
\citep{ServidioEA14}
and sharp gradients \citep{MarkovskiEA06}, and these may contribute to dissipation. 
 More work will have to be done to investigate, 
confirm and refine any of these emerging scenarios, in particular with respect
to the role of non-Maxwellian velocity space features 
seen in fully kinetic nonlinear simulations 
\citep{Hellinger13, ServidioEA12,ServidioEA14}.

\subsection{Magnetic reconnection, current sheets and intermittency}\label{sec:recon}

Magnetic reconnection is an important element 
of the dynamics of plasmas, and has been widely studied
for more than half a century both theoretically and in applications. 
The basic theory has been well reviewed in terms of MHD 
theory \citep{biskamp00,ForbesPriest,zweibel09}, 
the complications that emerge in three dimensions \citep{PriestSchrijver99},  
the transition from MHD to kinetic behavior
\citep{BirnPriest}, and in situ spacecraft observations
\citep{BirnPriest}, and 
plasma experiments where kinetic effects begin to become important 
\citep{YamadaEA10}. 
We will not attempt to reproduce such reviews here. 
Many of the computational models in which reconnection has been studied
(see references above) have been formulated to include kinetic
effects, but in small domains with 
simple boundary conditions and smooth initial data, 
often based upon an equilibrium current configuration.
In such cases the reconnection that is studied may be viewed as 
{\it spontaneous reconnection}, which itself is a rich and well-studied
approach. 
However it has become increasingly clear that
the traditional way of studying magnetic
reconnection via standard equilibrium 
current-sheet setups needs to be complemented by 
investigations of reconnection in more complex environments, 
and in particular in self-consistent turbulent environments. 
As of now, we are still in the early stages
of understanding this more complex situation, in particular with respect to 
kinetic treatments of quasi-collisionless systems. 
One key question in this 
context is the degree to 
which phenomena associated with reconnection, such as
particle acceleration,
can act as an alternative route to dissipation.
Here we can provide only a brief overview of this 
active research topic, which has been examined from a variety of
similar approaches. 

The first description of turbulent reconnection
\citep{MatthaeusLamkin86}, an initial value problem consisting 
of a sheet pinch evolving in the 
presence of a specified broad-band spectrum of fluctuations, might properly be called
``reconnection in the presence of turbulence.'' 
In this case one finds bursty, nonsteady reconnection, in which 
one observes sporadically forming
intense current sheets and vortex quadrupoles, as well as 
transient multiple X-points.  
This approach was found to lead to elevated rates of 
reconnection, for resistive MHD, the 
increase for large systems 
being comparable to or greater than the 
increase due to Hall effect \citep{SmithEA04}.
An important dynamical feature of this problem is the 
{\it amplification} of the turbulence due to nonlinear 
instability of the initial configuration and subsequent feedback
\citep{Lapenta08}. 
	
Another approach, which has usually been applied in three dimensions,
also begins with a sheet pinch initial condition,
but instead of supplying turbulence through an initial spectrum 
of fluctuations, a random source of fluctuations acts continuously 
through a forcing function applied to the region surrounding and 
including the current sheet 
\citep[see][and references therein]{KowalEA09,LoureiroEA09,LazarianEA12}.
This too gives rise to strong turbulence effects, and 
as might be expected, the reconnection rate generally 
is tied closely to the imposed turbulence amplitude. 

Still another approach in understanding the relationship between 
turbulence and reconnection is to initialize a system
with a large number of magnetic flux tubes, as well as random
velocity fields, such that the initial state triggers a complex cascade.
The turbulent dynamics leads to 
interactions between various pairs of 
adjoining magnetic flux tubes (or magnetic islands), 
leading to reconnection with widely distributed 
reconnection rates, studied in 2D MHD  \citep{ServidioEA09,ServidioEA10}
and more recently in Reduced MHD \citep[see below and][]{WanEA14}.
This type of ``reconnection in turbulence''
might be viewed as similar to both 
problems described above, but with the random perturbations
caused by the cascade itself, rather than being imposed 
by an initial spectrum
or by a prescribed forcing function. 
Some have argued that for 
systems having many flux tubes, this 
might be viewed as a more natural way
to drive reconnection with turbulence,
but it has the disadvantage of requiring a large 
system, both to 
establish a high Reynolds number cascade,
and to adequately resolve the smaller scale 
current sheets. 

A further complication in understanding reconnection 
is that its geometry can become quite complex 
in three dimensions \citep{PriestSchrijver99},
departing strongly from the familiar two dimensional 
forms. 
Nevertheless it seems 
rather certain that in 3D models, as in 2D models, 
coherent electric current structures, including
sometimes complex sheet-like structures \citep{MininniEA08}, 
continue to play an important role.   
For example, 
current sheets in RMHD 
models occupy a central role in 
models of coronal heating that have been 
studied \citep{EinaudiVelli99,DmitrukEA98,RappazzoEA10} 
in the so-called nanoflare scenario. This model
is typically viewed as an implementation of the
Parker problem \citep{Parker72} in which 
coronal field lines are stirred from below by photospheric 
motions, which causes a braiding or tangling of flux tubes, 
the formation of current sheets between pairs of them, and subsequent 
bursty reconnection and heating. 
Recently there has been further progress in understanding
the local statistics of 
current sheet dynamics in the weakly three dimensional 
Reduced MHD model discussed above, 
thus advancing out understanding of 
the role of these current sheets in 
reconnection, heating and intermittent dissipation in 3D. 
Other, simpler, models can be constructed that take into account the phenomenology 
of MHD by stipulating that dissipative structures are current and vorticity sheets 
and that the typical time of energy transfer to small scales is modified in MHD when 
taking into account the role of Alfv\'en waves
\citep[see, e.g.][]{graue94,Politano}.

\cite{ZhdankinEA13} carried 
out a quantitative statistical analysis of 
current sheets that emerge in RMHD turbulence,
and reported on the distribution functions of current 
sheets with respect to their dimensions, peak current densities, 
energy dissipation rates and other characteristics.
\cite{WanEA14}
reported a similar study using an RMHD 
coronal heating model, confirming many of
the results in Zhdankin et al.\ (2013), 
while also computing the distribution of reconnection rates
and demonstrating the statistical 
connection between current sheet 
dimensions and characteristic turbulence 
length scales. 
An interesting result obtained in both the above studies
is that the locus of maximum dissipation rate,
always the peak of current density for scalar 
resistivity, is {\it not} always, or even usually,
located at the component X-points. 
This is a property also 
found in laminar asymmetric reconnection
\citep{CassakShay07}, 
having different magnetic field strength 
on the two flow sides of the reconnection zone.   
Perhaps not surprisingly, 
larger reconnection rates are well correlated 
with the proximity of the current maximum to the 
X-type critical point.
These purely spatial analyses were extended into
the temporal domain by \cite{ZhdankinEA15}
who tracked the evolution of dissipative structures over
time and measured the statistics of their lifetimes and
total energy dissipation.  The results obtained
so far for reconnection in RMHD are clearly 
valuable
in providing some information about the 3D case,
and are specifically applicable to low plasma beta,
highly anisotropic systems driven at low frequencies,
such as the coronal flux tube problem. 
However the general 3D case will undoubtedly contain 
significantly greater complexity
\cite[see e.g.][]{MininniEA08}, 
much of which remains to be explored.

The dynamics of the 
formation of current sheets and other small scale coherent structures
is of great importance in understanding the 
intermittent cascade and its fate.
Furthermore, current sheet formation
may be quite different in a large turbulent system than it 
is in a laboratory device in which the magnetic field 
to leading order is large scale, laminar,
and controlled by external coils. 
For example, it is well known that 
ideal-MHD flows that develop in 
turbulence give rise to intense thin current-sheet 
structures \citep{FrischEA83,WanEA13}.
The ideal process of current sheet generation,
observed at short times in high resolution MHD simulations
\citep{WanEA13}, apparently gives essentially identical 
higher order magnetic increment statistics as are seen in comparable 
high Reynolds number simulations -- so we can understand that 
intermittency and the drivers of the conditions that 
lead to reconnection are ideal processes. 
In retrospect, this could 
have been anticipated in Parker's original discussion
of coronal flux tube interactions \citep{Parker72}.  
Reconnection may subsequently be triggered at these sites, 
resulting in dissipation of turbulent magnetic field.

The process of current sheet formation
in the presence of weak dissipation 
can also involve 
multiple magnetic X-points and secondary 
islands or flux tubes.  This was
originally observed in 
reconnection with finite background turbulence,
and in that context
it was suggested that secondary islands
might elevate the reconnection rate
 \citep{MatthaeusLamkin85,MatthaeusLamkin86,LoureiroEA09}.
\cite{Biskamp86} suggested that secondary islands might 
emerge due to a linear instability
of thin current sheets 
above magnetic Reynolds numbers of
about $10^4$.
 More recently this instability was revisited
based on the recognition that the tearing instability can become much
faster if it originates in a current sheet already thinned to 
the Sweet-Parker thickness \citep{LoureiroEA07,BhattacharjeeEA09}.
The occurrence of this ``plasmoid instability''
has been supported by simulation studies using 
MHD and PIC codes
\citep{SamtaneyEA09,DaughtonEA11b,LoureiroEA12}.
\cite{LoureiroEA07} found that the growth rate of the secondary tearing instability
of a Sweet-Parker reconnection layer is higher than the inverse global Alfven
transit time along the layer.  M.\ Velli and collaborators 
\citep[e.g.][]{PucciVelli14}
have argued that this result means that, in reality, such a layer cannot be formed
in the first place.
In fact when the the linear growth rate of tearing instability equals the 
inverse of $\tau_A=L/V_a$ the current sheet necessarily is disrupted.
This occurs at $a/L \sim S^{-1/3}$, where $S$ is the 
global Lundquist number, $S = L V_A/\eta$.  
In the asymptotic limit $S \rightarrow \infty$, this value is 
much greater than for the SP layer, which suggests that as a 
current layer is being formed, it is 
disrupted by secondary tearing well before it reaches the SP stage. 
A similar conclusion was presented by 
\citep{UzdenskyLoureiro14}.

As mentioned above, one also finds the  
occurrence of numerous secondary islands 
in a turbulence context, and this will also be the fate of any multiple 
island scenario at finite amplitude.
High resolution 2D MHD turbulence simulations display a proliferation of magnetic X-points
at sufficiently high magnetic Reynolds number $R_m$ \citep{WanEA12}, 
with the number of X-points and flux tubes observed in the simulations, 
scaling as $R_m^{3/2}$.
A cautionary word is that secondary islands (or plasmoids)
also form due to numerical error, and there is a requirement for 
careful resolution studies \citep{WanEA10}
to ensure that complex multiple plasmoid reconnection 
is physical and not numerical. 
It is unclear at present what precise 
relationship exists between plasmoid instability and 
generation of secondary islands turbulence. 
It is noteworthy that simulations, even laminar cases, 
usually trigger reconnection with a finite amplitude perturbation, 
and by the time multiple islands are observed there are many finite 
amplitude modes participating.
Whether the origin is instability or cascade, it seems  
certain that at high magnetic Reynolds numbers, reconnection 
zones will be complex, even in 2D, 
so that the details of the many individual 
reconnection processes will almost certainly not be resolved in LES,
but rather their aggregate effect will need to be built into a 
SGS model.
  
To close this section we note first 
the implications of the evolving perspectives on reconnection 
for an application, say, solar coronal heating. 
In such a complex driven system 
that is far from equilibrium, 
one should properly view 
current sheet formation, dissipation, magnetic reconnection, 
and nanoflares, not as independent processes, 
but rather as outcomes of a nonlinear MHD-turbulent cascade 
in a self-organized solar corona. Building such effects into an LES/SGS model will be challenging.

Finally for completeness we list some of the outstanding 
questions that emerge from this discussion:
\begin{enumerate}
\item Do singular (ideal) structures matter for the dissipative case?
\item Does the 2D case matter to understand the 3D case? 
\item Are rotational discontinuities a central piece of 3D reconnection? 
\item Does current sheet roll-up play a role? 
\item What role do invariants (magnetic and cross helicity) have in reconnection? 
\item Is the rate of dissipation independent of Reynolds number? 
\item What are the dissipative and/or reconnecting structures, and how to identify them
in a real (natural) system? 
\item What is the role of the magnetic Prandtl number? 
\item Do small-scale kinetic effects that emerge in reconnection
alter large-scale dynamics? How? 
\item Can Adaptive Mesh Refinement help in a general approach?
\end{enumerate}

\section{LES in MHD}\label{sec:lesmhd}

The discussion in the previous section (\S\ref{sec:small}) highlights 
some of the challenges in devising reliable SGS models for LES of 
MHD turbulence.  In this section we describe in more detail the 
practical implementation of LES/SGS modeling, focusing on explicit
approaches that employ formal filtering operations designed to decompose
the flow into large and small-scale components.  The large-scale 
motion is computed by solving the filtered non-stationary equations 
of MHD while the SGS terms are parameterized and expressed in terms 
of the filtered quantities.  Though real plasma flows are likely
much more subtle (\S\ref{sec:small}), current MHD-LES models
often assume that the subgrid scales (SGS), also referred to
as subfilter-scales (SFS), are relatively isotropic, homogeneous,
and universal.

LES is a method for simulation of flows with large Reynolds numbers.  
It is generally not valid for low Reynolds number flows since it assumes
that there is a substantial (order unity) nonlinear transfer to small 
scales.  Furthermore, as discussed in \S\ref{sec:small}, the usual 
assumption of isotropy at small scales may not be realized.  This
may occur in rotating and/or stratified flows if the cut-off 
wavenumber where the filter is applied is in the anisotropic 
range.  And, it may occur more generally in turbulent MHD flows 
where anisotropy may progressively increase toward smaller scales
unless this is mitigated by turbulent reconnection processes which
may help recover isotropy.

Initially the Large Eddy Simulation technique was developed for
the simulation of HD turbulence of neutral fluids,
particularly in the context of atmospheric and engineering applications
\citep{meneveau, sagaut06,Glazunov}.  This has been extended
to MHD turbulence by several authors who adapted known HD
closures to the MHD case and developed new SGS models. 
\citep{Agullo,Muller2,Muller3,Yoshizawa,Zhou,Knaepen}.  Our discussion
of the general approach follows that given in \citet{Chernyshov1}.
Our intention is to illustrate how the MHD equations can be cast into a a traditional LES framework.  We make no attempt at a comprehensive survey of existing LES/SGS models.  For alternative approaches see \cite{miki08}, \cite{grete15}, and \cite{schmi15}.   Of particular note is the model presented by \cite{grete15} in which the SGS stress tensor involves nonlinear correlations between the resolved (filtered) velocity and magnetic field gradients, along with Smagorinsky-like expressions for the SGS kinetic and magnetic energy proportional to the corresponding rate of strain tensors.  This is based on a Taylor expansion of the velocity and magnetic field within a filter box as originally proposed for HD by \cite{woodw06}.

As stated above, a LES applies a filtration operation 
to the primitive equations as suggested by
\citet{Leonard}\footnote{For an innovative filtering approach based on
wavelet transforms and adaptive mesh refinement see \citet{deste13}}.  
For the incompressible MHD equations,
the filter $G$ satisfies the following normalization
property:

%\AGP{{\underline{NOTE, AP:}} Here, Arakel has preferred and acute accent to a prime subscript; I have left this notation, not to confuse with the prime notion below for fluctuating fields}

\begin{equation}
      \int_{a}^{b} G (x_{j} - \acute{x_{j}},
  \bar{\triangle}_{j})d\acute{x_{j}} = 1.
\end{equation}

Here $G (x_{j} - \acute{x_{j}}, \bar{\triangle}_{j})$  is the
filter itself, of width $\Delta_j$.
Then, the filtered velocity is expressed as follows:

\begin{equation}
  \bar{u}_{j}=\int_{a}^{b} u(\acute{x_{j}}) G (x_{j} - \acute{x_{j}},
  \bar{\triangle}_{j})d\acute{x_{j}},
\end{equation}
where  $a=x_{j}-\frac{1}{2}\bar{\triangle}_{j}$ and $b
=x_{j}+\frac{1}{2}\bar{\triangle}_{j}$, $\bar{\triangle}_{j} =
(\bar{\triangle}_{x},\bar{\triangle}_{y}, \bar{\triangle}_{z})$.
$x_{j} = (x,y,z)$ are axes of Cartesian coordinate system.
The other physical fields are filtered similarly.

Let us present all the variables of the problem as the sum of
a filtered (large scale) and unfiltered (small scale)
component: $u = \bar{u}+u'$, $B = \bar{B}+B'$, $p =
\bar{p}+p'$ etc.,
with $u_j,\ B_j$ the velocity and  magnetic induction components, 
and $p$ the pressure.

To simplify the modeled equations describing compressible MHD flows, 
it is convenient to use mass-weighted filtering (also known as Favre
filtering) so as to avoid the appearance of additional SGS terms.
It is determined as follows:
\begin{equation}
   \tilde{f} = \frac{\overline{{\rho} f}} {\bar{\rho}} \label{E:favre}, \\
\end{equation}
with $\rho$ the density. The overline in eq.\ (\ref{E:favre}), 
denotes ordinary filtering while the tilde denotes mass-weighted 
filtering.  Mass-weighted filtering is used for all physical 
variables other than the pressure, density, and magnetic fields.

The Favre filtered velocity takes the following form
\begin{equation}
  \tilde{u}_{j}=\frac{\overline{{\rho} u_{j}}} {\bar{\rho}} =\frac {\int_{a}^{b} \rho u_{j}G (x_{j} - \acute{x_{j}},
  \bar{\triangle}_{j})d\acute{x_{j}}} {\int_{a}^{b} \rho(\acute{x}_{j}) G (x_{j} - \acute{x_{j}},
  \bar{\triangle}_{j})d\acute{x}_{j}}.
\end{equation}
The Favre filtered quantities can be presented in the form
of a sum, for instance for the velocity:  $u =\tilde{u} + u''$,
where the double prime designates the small-scale component.

The Favre-filtered MHD equations take the following form
\citep{Chernyshov1}:

\begin{itemize}
\item 
\emph{filtered continuity equation }

\begin {equation}\label{E:fneraz1}
  \frac{\partial{\bar{\rho}}}{\partial t} + \frac{\partial{\bar{\rho}} \tilde{u}_{j}}{\partial x_{j}} =
  0;
\end {equation}

%-------------------------------------------
\item 
\emph{filtered momentum conservation equation }

\begin{equation}\label{E:fNS1}
  \frac{\partial{\bar{\rho}} \tilde{u}_{i}}{\partial t} + \frac{\partial}{\partial
  x_{j}}\left (\bar{\rho} \tilde{u}_{i} \tilde{u}_{j} + \bar{p} \delta_{ij}
  -\frac{1}{Re}~\tilde{\sigma}_{ij} + \frac{\bar{B^{2}}}{2M_{a}^{2}}\delta_{ij} - \frac
{1}{2M^{2}_{a}}~\bar{B}_{j} \bar{B}_{i}
  \right) =\\
  =- \frac {\partial \tau_{ji}^{u}}{\partial x_{j}};
\end{equation}

%-----------------------------------------
\item 
\emph{filtered induction equation }

\begin{equation}\label{E:fin1}
  \frac{\partial \bar{B}_{i}}{\partial t} + \frac {\partial}{\partial
  x_{j}}\left ( \tilde{u}_{j} \bar{B}_{i} -
  \tilde{u}_{i}\bar{B}_{j} \right )- \frac{1}{R_m} \frac{\partial^{2} \bar{B}}{\partial x_{j}^{2}} =
   - \frac {\partial \tau_{ji}^{b}}{\partial
  x_{j}},
\end{equation}
%--------------------------------------------------------------------------------

as
\[
      \overline{\eta B_{j}}-\bar{\eta}\bar{B}_{j} = 0,
\]
\[
     \bar{\sigma}_{ij} - \tilde{\sigma}_{ij} = 0,
\]

\end{itemize}

\begin{itemize}
\item 
where
\[
   \tilde{\sigma}_{ij}= 2\tilde{\mu}\tilde{S}_{ij} - \frac{2}{3}\tilde{\mu}\tilde{S}_{kk}\delta_{ij} +
  \tilde{\zeta}\tilde{S}_{kk}\delta_{ij},
\]
\[
   \bar{\sigma}_{ij}= 2\overline{\mu S_{ij}} - \frac{2}{3}\overline{\mu S_{kk}\delta_{ij}} +
  \overline{\zeta S_{kk}\delta_{ij}},
\]
\item
and $\rho$ - density;~~ $p$ - pressure;~~ $u_{j}$ - velocity
in direction $x_{j}$;

$\sigma_{ij} = 2\mu S_{ij} - \frac{2}{3}\mu S_{kk}\delta_{ij} +
\zeta S_{kk} \delta_{ij}$  - viscous stress tensor;

$S_{ij} = 1/2\left( \partial u_{i} / \partial x_{j}
 + \partial u_{j}/ \partial x_{i} \right)$ - strain rate tensor;

$\mu$  - dynamic viscosity;~~$\zeta$ - bulk  viscosity;

$\delta_{ij}$ - the Kronecker delta;~~ $\varepsilon_{ijk}$ - the
Levi-Civita symbol;

$\eta=c^{2} / 4\pi\sigma$ - magnetic diffusion; $\sigma$ -
specific electric conductivity;

$F_{l} = \varepsilon_{ijk}j_{j}B_{k} / c$ - Lorentz force; $B$  -
magnetic field; $j$ - current density.
\end{itemize}

 $Re = \rho_{0}u_{0}L_{0} / \mu_{0}$ is the Reynolds number,
$R_m = u_{0}L_{0} / \eta_{0}$ - the magnetic Reynolds number. $
M_{s} = u_{0} / c_{s}$ the Mach number, $M_{a} = u_{0} / u_{a}$  -
the magnetic Mach number, and $c_{s}$ is sound speed determined by
the relation:
$c_{s}= \sqrt{\gamma p_{0} / \rho_{0}}$; $u_{a}$ is the
Alfv\'en speed, $u_{a} = B_{0}/ \sqrt{4\pi\rho_{0}}$.
The bulk coefficient of viscosity $\zeta$ is neglected.

The terms on the the right-hand-side of equations
~(\ref{E:fNS1})~-~(\ref{E:fin1}) designate the influence of
the SGS terms on the filtered component:

\begin{equation}\label{tauiju}
\tau_{ij}^{u} = \bar{\rho} \left ( \widetilde{u_{i} u_{j}} -
\tilde{u}_{i}\tilde{u}_{j}\right)- \frac {1}{M_{a}^{2}}\left(
\overline{B_{i} B_{j}} - \bar{B}_{i}\bar{B}_{j}\right) ;
\end{equation}

\begin{equation}
\tau_{ij}^{b} =  \left ( \overline{u_{i} B_{j}} -
\tilde{u}_{i}\bar{B}_{j}\right)- \left( \overline{B_{i} u_{j}} -
\bar{B}_{i}\tilde{u}_{j}\right).
\end{equation}

Note that compressibility alters the form of the SGS stress 
tensor $\tau_{ij}^{u}$ but the magnetic SGS tensor $\tau_{ij}^{b}$ 
is the same as for incompressible MHD.  The SGS-scale hydrodynamic
pressure in typically neglected in the filtered equations for compressible,
neutral flows with low Mach numbers \citep{piomelli99,zang92}.
By extension, the SGS magnetic pressure is also neglected in
eqn.\ (\ref{tauiju}).  However, attempts are being made at more general models.  For example, \cite{grete15} incorporate the full SGS strain rate tensors, including the symmetric components that account for SGS kinetic and magnetic pressure.

Let us consider the filtered equations
~(\ref{E:fneraz1})--(\ref{E:fin1}) in more detail. As far as the
small-scale velocity (and the other flow variables)  
$u'' = u -\tilde{u}$ is unknown; it has to be estimated with 
the use of the large-scale velocity obtained
by means of filtration. Theoretically, there is no functional
dependence between the small-scale velocity $u''$ and the large-scale
one $\tilde{u}$, so any estimation of $u''$ will contain error. 
DNS can sometimes be used to estimate this error but this 
can only be carried out for relatively low Reynolds numbers  
due to limited computational resources.

Thus, the filtered system of MHD equations contains the 
unknown turbulent tensors $\tau_{ij}^{u}$ and $\tau_{ij}^{b}$. 
The task of the SGS model is to express these unknown
tensors in terms of the filtered flow components
$\tilde{u}_{i}$  and  $\bar{B}_{i}$ using some sort 
of turbulent closure (parameterizations).   Ideally,
the closure model should capture such effects as
the Richardson turbulent cascade.

Let us consider closures for $\tau_{ij}^{u}$ and
$\tau_{ij}^{b}$. To guarantee the non-negativity of subgrid
energy, these tensors must satisfy some conditions, called 
realizability conditions.  A necessary and sufficient condition of
non-negativity is provided by the positiveness of the semidefinite
form for the turbulent tensors $\tau_{ij}$ such that:

\begin{equation}\label{E:realiz}
\begin{array}{c}
{\tau_{ii} \geq 0 ~~~ \mbox{for} ~~~i\in \{1,2,3\} }  \ , \\
{|\tau_{ij}| \leq \sqrt{\tau_{ii}\tau_{jj}}~~~ \mbox{for} ~~~ i,j\in
\{1,2,3\}}  \ , \\
{\mbox{det} (\tau_{ij}) \geq 0} \ .
\end{array}
\end{equation}

Let us assume that the form of the turbulent tensor $\tau_{ij}^{u}$  is
analogous to the viscous stress tensor (eddy viscosity
model), while $\tau_{ij}^{b}$ is analogous to ohmic dissipation.
This yields:
\begin{equation}\label{E:tau_u}
\tau_{ij}^{u} - \frac{1}{3} \tau_{kk}^{u}\delta_{ij} = -2\nu_{t}
\left ( \tilde{S}_{ij} - \frac{1}{3} \tilde{S}_{kk} \delta_{ij}
\right),
\end{equation}
%-------------------------------------------------------
\begin{equation}\label{E:tau_b}
\tau_{ij}^{b} - \frac{1}{3} \tau_{kk}^{b}\delta_{ij} = -2\eta_{t}
 \bar{J}_{ij},
\end{equation}

 where
   \[
    \tilde{S}_{ij} = \frac{1}{2}\left( \frac{\partial \tilde {u}_{i}}{\partial
x_{j}} +\frac{\partial \tilde{u}_{j}}{\partial x_{i}}\right)  \]\\
is the large-scale strain rate tensor;

    \[
    \bar{J}_{ij}= \frac{1}{2}\left( \frac{\partial \bar
{B}_{i}}{\partial x_{j}} - \frac{\partial \bar{B}_{j}}{\partial
x_{i}}\right)  \] \\  is a large-scale magnetic rotation tensor.
Here $\nu_{t}$ and $\eta_{t}$ are scalar turbulent diffusion 
coefficients that may in general depend on the spatial coordinates 
and time.

In the right-hand side of equations ~(\ref{E:tau_u}) and
~(\ref{E:tau_b}) the symmetric terms of the magnetic
rate-of-strain tensor:  \[\bar{S}_{ij}^{b} = \left(\partial \bar
{B}_{i} / \partial x_{j} + \partial \bar{B}_{j}/\partial
x_{i}\right)/2,\] and vorticity tensor:\[ \tilde{J}_{ij}^{u} =
\left(
\partial \tilde{u}_{i} /
\partial x_{j} - \partial \tilde{u}_{j}/ \partial
x_{i}\right)/2\] are omitted, because their
contribution is negligible in many circumstances \citep{Muller2}. 
However, \cite{grete15} have incorporated these symmetric terms and find that they may be important in particular for supersonic flow regimes, as encountered, for example, in the interstellar medium.
We note again that 
the main purpose of SGS modeling is not to
fully reconstruct the information lost due to filtration but 
rather to accurately capture the influence of the SGS flow
on the large-scale energy distribution and transport.

Often the term  $\frac{1}{3} \tau_{kk}^{u} \delta_{ij}$ is
combined with the thermodynamic pressure, 
$\nabla (p + \frac{2}{3}k \delta_{ij})$, where
$k = (\tau_{11} + \tau_{22}+ \tau_{33})/2$ is the SGS 
turbulent kinetic energy \citep{Erlebacher}. 
In the present paper we consider the isotropic
component explicitly, though the isotropic component
of the magnetic tensor ~(\ref{E:tau_b}) vanishes
because of vanishing $J_{ii}$).   The isotropic 
component of $\tau^u$ can be found from the realizability
conditions ~(\ref{E:realiz}), which give
\begin{equation}
\tau_{12}^{2} + \tau_{13}^{2} + \tau_{23}^{2} \leq
\tau_{11}\tau_{22} + \tau_{11}\tau_{33} + \tau_{22}\tau_{33}
\end{equation}

By using ~(\ref{E:tau_u}), we obtain the following
expression for the isotropic component of 
$\tau^u$
\begin{equation}\label{E:kinet}
k \geq \frac{1}{2}\sqrt{3}( \nu_{t}|S^{u}| ) \ ,
\end{equation}
where 
$|\tilde{S}^{u}|=\left(2\tilde{S}_{ij}^{u}\tilde{S}_{ij}^{u}\right)^{1/2}$.
The anisotropic and isotropic components of $\tau^u$ can then
be obtained from Eqs.\ (\ref{E:tau_u}) and (\ref{E:kinet}).

Different closures for the compressible MHD equations
were developed by \cite{Chernyshov2} and further analyzed
by \cite{Chernyshov4}.  LES of MHD turbulence were compared 
with DNS and it was shown that the five closure models considered 
provide sufficient dissipation of kinetic and magnetic energy 
at comparatively low computational expense.

The effects of heat conduction were considered by
\cite{Chernyshov3,Chernyshov5} who developed models for 
the SGS terms in the energy equation as well as for the 
magnetic terms in the momentum and induction
equations.  LES of decaying MHD turbulence were performed and
their greater efficiency compared to DNS was demonstrated.
SGS models were similarly validated for studies of self-similar
regimes in forced turbulence by \cite{Chernyshov8,Chernyshov9}.
Further work on the LES of compressible MHD turbulence focused on
the local interstellar medium \cite{Chernyshov6} and the kurtosis 
and flatness of the turbulent flow \cite{Chernyshov7}.
For a comprehensive review of SGS modeling see \cite{Chernyshov10}.

%%%%%%%%%%%%%%%%%%%%%%%%%%%%%%%%%%%%%%%%%%%%%%%%%%%%%%%%%%%

Other models can be developed, using for example expressions deriving from the integro-differential equations for energy spectral density in the framework of  two-point closures of turbulence like the Eddy Damped Quasi Normal Markovian closure  \citep{chollet}.
Newer versions have seen several developments implemented, such as (see \cite{ baerenzung11} and references therein):
(i)  including both eddy diffusivities (viscosity and resistivity) as well as eddy noise; indeed, the effect of the small scales on the large scales is potentially a dissipation of energy (although eddy diffusivities can be negative) as well as  a stochastic forcing;
(ii)  adapting to spectra that differ from the classical Kolmogorov law, a feature that can be useful in the presence of magnetic fields, rotation or stratification, i.e. when the resulting energy spectra may be in a weak turbulence regime due to wave-eddy interactions; and
(iii) including in these two types of coefficients  the effect of helicity (velocity-vorticity correlations) as encountered in tropical cyclones or in the Planetary Boundary Layer. Helicity can be created by a combination of rotation and stratification and can play a role in the generation of large-scale magnetic fields.

In fact, in MHD, there are two other helical fields that can be defined, namely the cross-correlation between the velocity and the magnetic induction, and the magnetic helicity (correlation between vector potential and magnetic induction in three space dimensions); their effect on the large-scale dynamics of MHD flows has been considered in  \cite{yokoi_13} where the role of cross-correlation on turbulent reconnection is particularly stressed \cite[see also][]{yokoi_93}.

There are other types of LES that have been developed.  Of particular note are implicit LES (ILES) methods that do not include any explicit SGS model but do include intrinsic dissipation and dispersion due to the nature of the numerical algorithm.  These methods received much attention at the workshop and make up a growing fraction of astrophysical and geophysical turbulence simulation.  Their primary attraction is maximal resolution; dissipation operates only at the grid scale, leaving larger scales essentially free of artificial diffusion.  However, it can be difficult to assess the influence of the dissipation scheme on the properties of the resolved flow, particularly in the case of MHD where nonlinear spectral interactions are intrinsically nonlocal.  For further details on ILES see \cite{ILES} and \cite{schmi15}.

Another possibility that has been tested in two dimensions in MHD in a pseudo-spectral code is to decimate Fourier modes after a given cut-off scale \citep{meneguzzi}; the rationale behind such a decimation, whereby for example half the modes are taken for Fourier shells beyond the chosen cut-off (and the method can be iterated),  is that there are $4\pi k^2$ modes of characteristic wavenumber k, i.e. a large number at small scales; it is to be expected that their role is statistical (and with a stochastic component) and that therefore these modes do not have to be all treated explicitly. For example, in a computation with $3072^3$ grid points and with a de-aliasing using a 2/3 rule, the ratio of maximum to minimum wave numbers is $1024$. Of the $27^+ \times 10^9$  modes in such a computation, half of them (or roughly $13$ billions) are for wave numbers $k\ge 710$, with $\approx 12 \times 10^6$ in the very last Fourier shell (of unit width) alone \citep{marino13}.

In conclusion, it has been suggested that a possible future methodology to tackle complex turbulent flows as found in astrophysics and space physics might  be to combine multiple approaches, including ILES, explicit SGS modeling, and also adaptive mesh refinement \citep{woodw06,deste13,schmi15}

\section{Applications}\label{sec:applications}

In this section we briefly consider several applications of LES in MHD of relevance to 
astrophysics and space physics in order to highlight both the successes and the challenges
of the field.

%AB: written this about solar surface modeling, which we discussed at the meeting
\subsection{Realistic LES of solar granulation}\label{sec:solar_granulation}

A prime example demonstrating the success of LES is solar granulation.
This includes in particular the uppermost visible surface of the solar
convection zone where radiation transport and time-dependent ionization
are important.
Following the early work \citep{Spi71,Spi72,GMSW76,TZLS76} on compressible
stellar convection in the 1970s, \cite{Nor82,Nor85} pioneered the field of
realistic convection simulations of the solar surface, which has since
advanced considerably \citep{SN89,SN98,SLK89,WFSLH04,VSSCE05,rempe14}.
These simulations employ a tabulated equation of state together with fully
nonlocal radiation transfer and realistic opacities.
They used different combinations of subgrid scale modeling including
Smagorinsky viscosity, shock-capturing viscosities, hyperviscosities,
Riemann solvers, monotonicity schemes, etc., which can be classified
as implicit LES (ILES).

\begin{figure}[t!]\begin{center}
\includegraphics[width=.98\textwidth]{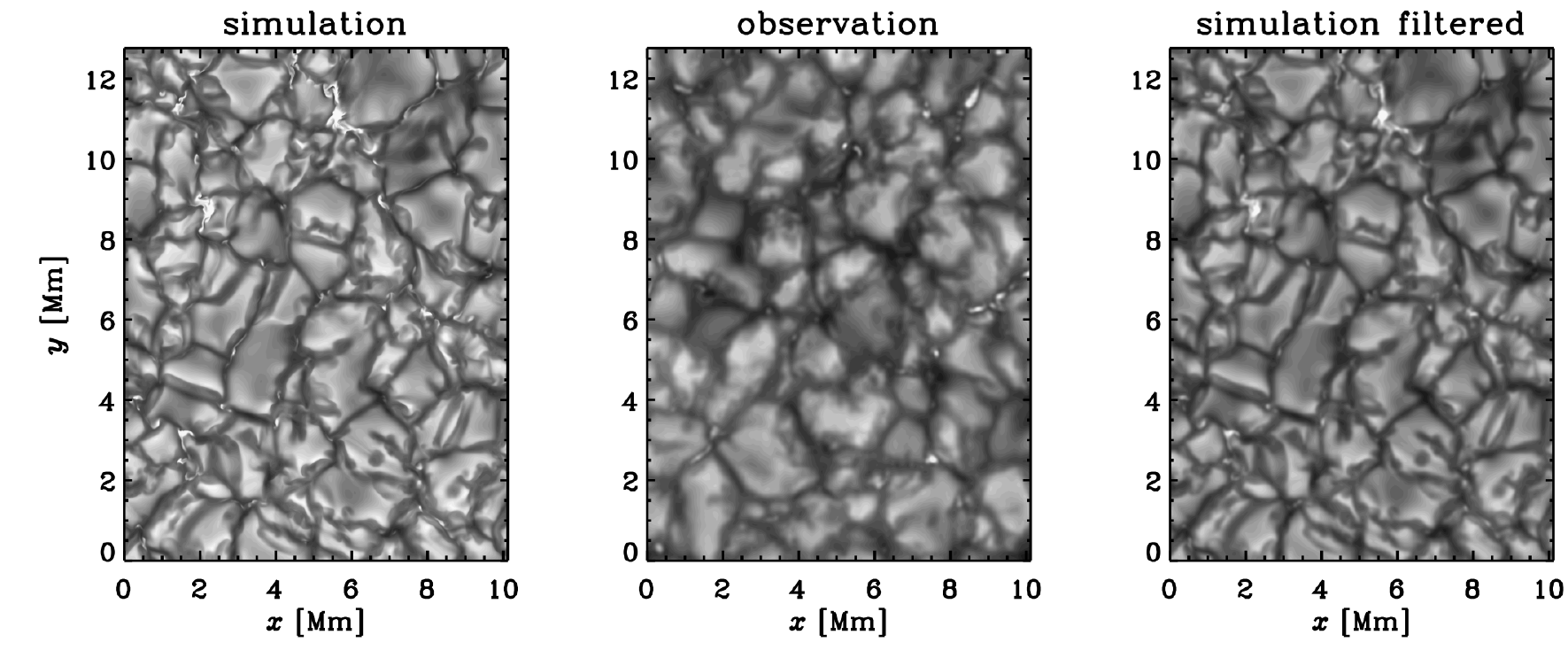}
\end{center}\caption[]{
Comparison between a granulation pattern from a simulation with 12 km
grid size (left), an observed granulation pattern from the
Swedish 1-meter Solar Telescope at disk center
(middle), and the simulated one after convolving with
the theoretical point spread function of a 1 meter telescope.
The simulation images are for wavelength integrated light intensity
while the observed image is for a wavelength band in the near UV.
The image was taken on 23 May 2010 at 12:42 GMT with
image restoration by use of the multi-frame blind de-convolution
technique with multiple objects and phase diversity \citep{vNetal05}.
Courtesy of V.~M.~J.~Henriques and G.~B.~Scharmer and adapted from \cite{BN11}.
}\label{GranulationFig}\end{figure}

Simulations of solar surface convection reproduce solar observations
remarkably well, both qualitatively and quantitatively.  An example
is shown in Fig.\ \ref{GranulationFig}.  The intensity contrast in the 
simulated granules is about 16\%, which agrees with the observed contrast
of about 10\% after taking atmospheric seeing and the telescope point spread
function into account \citep{SN98}.  Other quantitative successes include
power spectra, spectral line formation, acoustic mode excitation, and
local dynamo action \citep{nordl09,rempe14}.

An important question is to what extent the success of these simulations
is due to the ILES technique employed, or to aspects of the physics
that make this application particularly amenable to LES modeling.
For example, the strong density stratification makes convection
highly anisotropic, with a dilution of vorticity tending to 
``laminarize'' upflows while the dynamics of the downflows are
controlled mainly by buoyancy and entrainment.  So, details such as 
the forward transfer of kinetic energy into the dissipation scale
are not specifically tested.  Furthermore, near the visible 
surface, the radiative diffusivity is large enough that no SGS model
is needed for the internal energy equation.  Thus, a key component
of the dynamics is effectively captured through DNS.  This may also
account for the success of geodynamo models, which are able to
run with a realistic value for the magnetic diffusivity, relegating
SGS to the velocity field alone \citep{glatz02,jones11}.

\subsection{The bottleneck effect in HD turbulence}\label{sec:bottleneck}

\begin{figure}\centering\includegraphics[width=.8\textwidth]{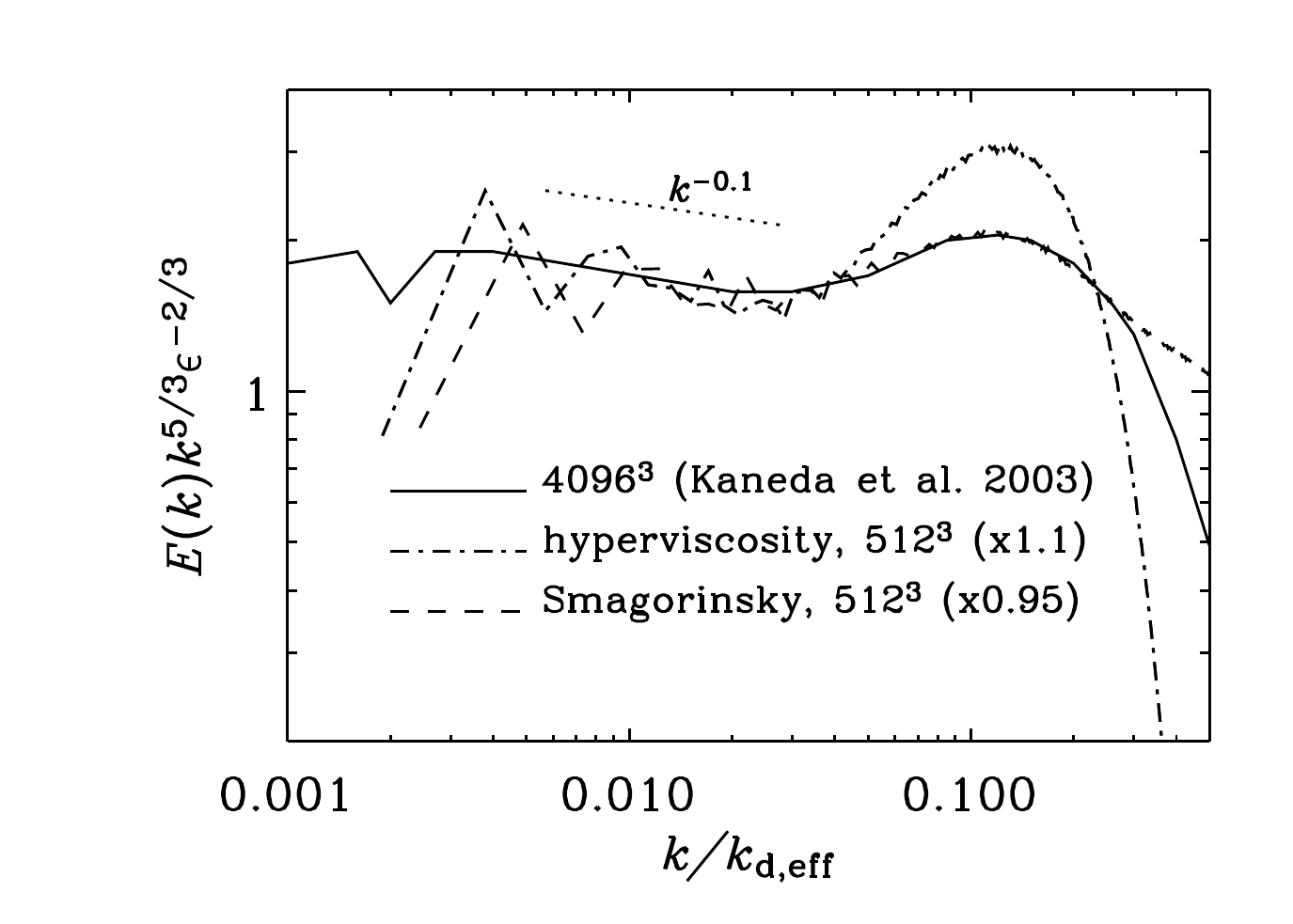}
\caption{Comparison of energy spectra of the $4096^3$ meshpoints run \citep{KIYIU03},
solid line, and $512^3$ meshpoints runs with
hyperviscosity (dash-dotted line) and Smagorinsky viscosity (dashed line).
(In the hyperviscous simulation we use $\nu_3=5 \times 10^{-13}$.)
The Taylor microscale Reynolds number of the Kaneda simulation is 1201,
while the hyperviscous simulation of \cite{HB04} has an approximate
Taylor microscale Reynolds number of $340<\mbox{Re}_\lambda<730$.
For the Smagorinsky simulation the value of $\mbox{Re}_\lambda$ is
slightly smaller.
Courtesy of Nils E.\ Haugen \citep{HB06}.
\label{kan_hyp_smag}}
\end{figure}

Incompressible forced turbulence simulations have been carried out at
resolutions up to $4096^3$ meshpoints \citep{KIYIU03}.
A surprising result from this work is a strong bottleneck effect \citep{Fal94}
near the dissipative subrange, and possibly a strong inertial range correction of
about $k^{-0.1}$ to the usual $k^{-5/3}$ inertial range spectrum,
Interestingly, similarly strong inertial range corrections have also been seen
in simulations using a Smagorinsky subgrid scale model \citep{HB06}
using $512^3$ meshpoints; see the dashed line in Fig.\ \ref{kan_hyp_smag}.
Here we also show the results of simulations with hyperviscosity,
i.e.\ the $\nu\nabla^2$ diffusion operator has been replaced by a
$\nu_3\nabla^6$ operator \citep{HB04}, also with $512^3$ meshpoints
(dash-dotted line).  Hyperviscosity greatly exaggerates the bottleneck 
effect, but it does not seem to affect the inertial range significantly;
see Fig.\ \ref{kan_hyp_smag}.  
\cite{woodw06,woodw06b} found a similar bottleneck effect in ILES of homogeneous, 
decaying, sonic turbulence (Mach number 1).  Furthermore, they implemented 
a nonlinear SGS model (mentioned previously in \S\ref{sec:lesmhd}) 
designed to supplement the numerical dissipation and they found that 
the combined ILES+SGS model could both alleviate the bottleneck effect and 
reproduce the spectrum of higher-resolution ILES across much of the 
resolved dynamical range.

If the details of the inertial range spectrum are sensitive to the
dissipation even in this most fundamental of applications then 
what should we expect for more complex flows?  Does this call 
into question the {\it central premise} discussed in 
\S\ref{sec:intro}, that the dynamics of the large scales can be
reliably captured despite the challenges of modeling the 
SGS physics?

\subsection{Problems in dynamo theory}\label{sec:dynamo_theory}

\subsubsection{Small-scale dynamos}\label{sec:SSdynamos}

Unlike many industrial applications where LES have been tested against
experiments, this is currently impossible for hydromagnetic flows
exhibiting dynamo action.
Except for astrophysical dynamos, there are not even a hand-full of
laboratory experiments to date that produce self-excited hydromagnetic
dynamo action.
Therefore, an important benchmark is provided by DNS.

Dynamos come in two flavors: small-scale and large-scale dynamos
\citep[see][for recent reviews]{BS05,BSS12}.
In the kinematic phase, during which the magnetic field grows exponentially
from a weak seed magnetic field, the magnetic field exhibits a $k^{3/2}$
\cite{Kaz68} spectrum.
Evidently, this spectrum diverges toward small length scales, so one cannot
expect to obtain the correct growth rate with LES.
This has consequences for understanding the excitation conditions of
small-scale dynamos.
DNS have demonstrated that the onset of small-scale dynamo action
depends on the value of the magnetic Prandtl number
(see Iskakov et al.\ 2007 and Brandenburg 2011 for the classical incompressible 
case and Fedderrath et al.\ 2014 for the case of supersonic turbulence).
\nocite{iskak07,Bra11,feder14}
Meanwhile, the magnetic Prandtl number does not enter into traditional
LES, so this question cannot be addressed.
This is also the case for the magneto-rotational instability (MRI),
which has been shown to be sensitive to the value of the magnetic Prandtl number.

As the small-scale dynamo saturates, the peak of magnetic energy moves
gradually toward larger scales, so there is a chance that this can be
modeled with LES.
However, simulations using a Smagorinsky-like magnetic diffusivity prescription
have yielded saturation field strengths that are significantly below those
obtained with DNS \citep{HB06}.
This shortcoming might be related to not yet being in the asymptotic regime
in which {\it both} magnetic and kinetic Reynolds numbers are large enough.
A similar situation might apply to the ratio of kinetic to magnetic energy
dissipation, which is known to scale with magnetic Prandtl number to a power
that is around 0.6.
Again, LES are not currently able to shed any light on this, because the
magnetic Prandtl number does not enter in standard subgrid scale models.
Addressing this deficiency is a difficult but perhaps auspicious challenge 
in need of further research.

\subsubsection{Large-scale dynamos}\label{sec:LSdynamos}

Large-scale dynamos produce magnetic fields whose scale exceed that of the
turbulent eddies.
They are believed to be relevant for understanding the global 22-year cycle
of the Sun's magnetic field, and similar large-scale magnetic fields in
other astrophysical bodies.
A leading theory for understanding large-scale dynamos is mean-field
theory which explains the occurrence of correlation in the mean electromotive
force that has a component parallel to the mean magnetic field.
This is generally referred to as the $\alpha$ effect, is typically related
to the presence of helicity in the system.
However, DNS have shown that its magnitude is reduced with increasing values
of the microphysical magnetic Reynolds number \citep{CH96}.
This is now generally referred to as catastrophic quenching and has to do
with a magnetic contribution to the $\alpha$ effect, which is proportional
to the current helicity of the fluctuating field, $\overline{\jj\bdot\bb}$.
Here, $\jj=\nab\cross\bb/\mu_0$ is the current density of the fluctuating
magnetic field, $\bb$.
It is in turn related to the magnetic helicity of the small-scale field,
$\overline{\aaaa\cdot\bb}$, where $\aaaa$ is the magnetic vector potential
of $\bb=\nab\cross\aaaa$.
It obeys the evolution equation
\EQ
{\partial\over\partial t}\overline{\aaaa\cdot\bb}=
-\overline{\uu\cross\bb}\cdot\meanBB-2\eta\mu_0\overline{\jj\cdot\bb}
-\nab\cdot(\overline{\ee\cross\aaaa}),
\label{dabdt}
\EN
Here, $\overline{\uu\cross\bb}$ is the mean electromotive force,
$\eta$ is the microphysical magnetic diffusivity, and
$\overline{\ee\cross\aaaa}$ is the magnetic helicity flux,
where $\ee$ is the fluctuating electric field.
We shall now discuss two aspects of this equation that are
relevant to LES.

First, if the system is homogeneous, i.e., there are no boundaries
and no large-scale variations of turbulent intensity and helicity
across the system, the divergence of the magnetic helicity flux vanishes.
In that case, there must be a balance between the first two terms on
the right-hand side of \Eq{dabdt}.
Although the assumption of homogeneity is only of academic interest,
it does provide a test case that must be obeyed equally by DNS and LES.
In particular, replacing the microphysical diffusion by hyperdiffusion
changes the scale dependence of the $\overline{\jj\cdot\bb}$ term and
has been shown to exaggerate the amplitude of the large-scale field
relative to that of the small-scale field \citep{BS02}.
This phenomenon is well understood, but it would still be of interest to
experiment with other representations of small-scale magnetic dissipation
to see how the large-scale magnetic field is being artificially modified
by the numerical representation of the small-scale physics.

Secondly, in the inhomogeneous case, magnetic helicity fluxes are possible.
In principle, these fluxes are gauge-dependent, and would thus be unphysical.
If, in the statistically steady state, the time derivative of
$\overline{\aaaa\cdot\bb}$ on the left-hand side of \Eq{dabdt} vanishes,
we have
\EQ
0=-2\overline{\uu\cross\bb}\cdot\meanBB-2\eta\mu_0\overline{\jj\cdot\bb}
-\nab\cdot(\overline{\ee\cross\aaaa}),
\label{dabdt0}
\EN
which implies that $\nab\cdot(\overline{\ee\cross\aaaa})$ is now balanced
by terms that are manifestly gauge-independent.
This is a remarkable property that allows us to measure the magnetic
helicity flux divergence.
This has been done in several recent papers \citep{MCCTB10,HB10,DSGB13}.
Interestingly, it turns out that the $\nab\cdot(\overline{\ee\cross\aaaa})$
remains subdominant for all simulations performed so far, and that only
at the largest resolution available so far, it becomes approximately
equal to the $2\eta\mu_0\overline{\jj\cdot\bb}$ term.
Here, the magnetic Reynolds number based on the wavenumber of the energy-carrying
eddies is about 1000, which is still barely achievable in DNS.

It is at present unclear whether LES are able to lead to meaningful insight
into the regime of larger magnetic Reynolds numbers.
One practical difficulty in determining $\overline{\ee\cross\aaaa}$ is the
computation of the magnetic vector potential, which is not always
readily available.
%AB.

\subsection{Astrophysical Turbulence and MRI}\label{sec:MRI}

Most astrophysical plasmas are highly compressible.  Thus, capturing 
the influence of shocks is essential.  Since the viscous scale 
in shocks is far too small to be resolved, DNS is not possible so 
modelers must turn to LES.  All shock-capturing astrophysical 
codes include some sort of subgrid-scale model, whether it be an 
explicit artificial viscosity \citep[e.g.][]{stone92} or an
implicit numerical dissipation that arises through
a Riemann solver as in Godunov-type methods 
\citep[e.g.][]{Roe86,leveq02}. These methods have been used for 
over 50 years, and there is no question this approach works 
extremely well.

Developing more sophisticated SGS models for turbulence in
highly compressible astrophysical plasmas will be a formidable
challenge.  A complete understanding of how shocks and contact 
discontinuities interact with the turbulent cascade and
affect small scales is still lacking and is a major research
problem in its own right.

One very important example of astrophysical turbulence in which 
compressibility does not play an essential role is the magnetorotational
instability, or MRI \citep{balbu91}.  MRI-induced turbulence is 
thought to dominate the energy and angular momentum transport 
in magnetized accretion disks \citep{balbu98}.  

The MRI is a particularly important case study for LES of MHD turbulence
because the role of artificial dissipation has been closely scrutinized
in recent years.  This scrutiny began with an influential paper by
\cite{froma07} that demonstrated a decrease in the amplitude of the
MRI-induced turbulent stresses with increasing spatial resolution
for the specific case of a local shearing box with no net flux 
through the layer.  The potential implications were profound; If 
this trend were to continue to the dynamic ranges active in actual 
accretion disks, then the turbulent transport would be drastically 
less efficient than previously thought and insufficient to account 
for the outward angular momentum transport necessary to sustain
the accretion process \citep{balbu98}.   They attributed this behavior 
to the form of the numerical diffusion and its interaction with the 
source terms that sustain the instability.  Other LES simulations soon 
confirmed this result for similar model configurations and demonstrated 
that the decrease in turbulent stresses with increasing resolution does 
not occur when explicit diffusion is included 
\citep[see][and references therein]{bodo11}.

However, after nearly eight years of active research, it appears that this
``convergence problem'' is not as serious as initially thought; the problem
appears to be a symptom of this particular model setup and goes away
when other model configurations are considered.  For example, if 
the vertical extent of the computational domain is increased so that 
it is twice as large as the horizontal extent, the convergence
problem goes away, the turbulent stresses increase by
an order of magnitude, and the Fourier power spectrum is 
altered substantially.  Furthermore, the convergence problem 
does not arise for shearing boxes with a background density stratification
and/or a net magnetic flux through the layer \citep[e.g.][]{froma13,turne14}.

In the no-net-flux case, the MRI must sustain a shear dynamo
and it may be that the properties of this dynamo (which is a macroscopic flow 
related to the outer scale of the turbulence) is not properly captured in small 
boxes and this introduces an artificial dependence on SGS diffusion.
Further insight into why the MRI behaves so differently in small,
non-stratified, shearing boxes with no net flux requires a deeper 
understanding of the MRI-driven dynamo, possibly based on reduced models.
This deeper understanding is also needed to account for the magnetic
cycles found in the larger boxes \citep[see][]{froma13,turne14}.

Furthermore, direct measurement of turbulent resistivity in the MRI by three 
different groups all find the same result, using very different codes and Reynolds
numbers \citep{guan09,lesur09,froma09}.  The turbulent magnetic Prandtl number is close 
to one, which implies the resistivity is very large (of order $u \ell$, where 
$u$ and $\ell$ are characteristic turbulent velocity and length scales; 
this implies that the turbulent 
magnetic Reynolds number is much smaller than that given by the Ohmic resistivity).  
This argues that macroscopic (turbulent) effects are more important than microscopic 
diffusivities, which in turn argues that the MRI is not sensitive to SGS physics.

Further confidence in the ILES approach to modeling MRI turbulence comes from the 
recent study by \cite{meheu15}.  They compared lower-resolution ILES to high-resolution 
DNS and found good agreement, at least for the special case of a non-zero mean 
field and a low magnetic Prandtl number, meaning that the magnetic diffusion was 
captured explicitly while the kinetic energy dissipation was relegated to the 
numerical diffusion.  In particular, the kinetic and magnetic power 
spectra at low to intermediate wavenumbers in a DNS with resolution (800, 1600, 832)
were well reproduced by ILES runs with resolutions of order 128$^3$ and 256$^3$ 
at a fraction of the computational expense.

If, on the other hand, one wishes to construct explicit SGS models for LES of MRI, it may
prove beneficial to exploit the potential magnetic induction introduced by 
\cite{salhi12}, ${\bf B} \bdot \del \Theta$ where $\Theta$ is the potential temperature.
This is an analogue of the potential vorticity which, unlike the potential vorticity, 
is a Lagrangian invariant for a magnetized, Boussinesq fluid.

\subsection{Hybrid Kinetic-MHD Models}\label{sec:hybrid}

In \S\ref{sec:kinetic} we emphasized that the small-scale dynamics 
of a plasma flow are often not well represented by MHD.  This is particularly
the case for low-density plasmas such as the solar wind or Earth's magnetosphere.
Departures from MHD must be treated by solving the kinetic equations in some 
form, often with simplifying assumptions designed to mitigate the computational
requirements.  See \S\ref{sec:kinetic} for further details and for a survey
of some applications from solar and space physics.

Many other astrophysical flows require a kinetic-MHD description to capture the
essential physics.  Here there is no question that SGS physics is important.
For example, anisotropic conduction and viscosity on small scales can influence
the large-scale dynamics through the magneto-thermal and magneto-viscous
instabilities \citep{balbu00,balbu01,quata02}.  There is some hope
that hybrid PIC simulations of small scales will be able to provide a 
reasonable SGS model (including coefficients of conduction and viscosity) 
that can be used in kinetic MHD models of various astrophysical problems
including the MRI (see \S\ref{sec:MRI}), hot gas in galaxy clusters, and 
turbulence in the interstellar medium.

During the GTP workshop, W. Schmidt described a promising hierarchical approach 
for including small-scale kinetic reconnection effects as an SGS model 
in MHD simulations.  The approach is similar to self-consistent mean-field 
dynamo theory but encompasses three stages: 
\begin{enumerate}
\item kinetic simulations of reconnection and dissipation providing effective transport coefficients 
(e.g., the turbulent EMF $\alpha$ and $\beta$ coefficients) to account for kinetic plasma processes; 
\item In turn, these coefficients are to be used in non-ideal MHD simulations of turbulent dynamos and reconnection, providing a sub-grid model for stage 3; 
\item quasi-ideal simulations of MHD turbulence. 
\end{enumerate}
 
\section{Summary and Outlook}\label{sec:summary}

The diverse applications surveyed in section \ref{sec:applications} all have
at least two things in common.  First, these systems are well described 
on large scales by the equations of MHD and second, they are characterized
by turbulent parameter regimes that are inaccessible to DNS.  Computers
simply are not capable of modeling all relevant scales from the macroscopic
scales to the Larmor radius.  Thus, in order to model such systems it is 
absolutely essential that we adopt an LES approach.

For many systems it may be sufficient to simply minimize the artificial
dissipation through the use of numerical methods that include their own
intrinsic dissipation.  Such methods, often referred to generally as 
implicit LES \cite[though see][for a more precise definition of ILES]{ILES}, 
maximize a simulation's dynamic range for a given spatial resolution by 
confining the artificial dissipation to scales comparable to the 
grid spacing.

Other applications may be more subtle.  For these, it may be necessary to 
model the subgrid-scale physics more reliably in order to accurately
capture the dynamics of the larger scales.  This can be achieved by 
applying a formal filtering procedure to the governing equations
(Sec.\ \ref{sec:lesmhd}) and then introducing parameterized or 
tabulated SGS models based on theoretical and phenomenological 
arguments or on local simulations that capture the small-scale 
plasma physics.  

When devising SGS models for MHD, some guidance can be provided
by the much more mature field of LES/SGS modeling in turbulent
HD (non-magnetic) flows, which has received much 
attention particularly in the context of atmospheric and 
engineering applications \citep{sagaut06}, with a growing body 
of literature also on highly compressible astrophysical 
flows \citep{schmi15}.  Still, as discussed in sections 
\ref{sec:large}--\ref{sec:small}, MHD possesses its own unique
challenges.   

Some of the challenges in representing SGS physics arise from 
the nature of the  MHD equations themselves.  The presence of 
magnetism in a turbulent, electrically conducting fluid introduces 
an intrinsic anisotropy to the flow that becomes more pronounced 
with decreasing scale (\S\ref{sec:small}).  The small-scale flow 
is also intrinsically inhomogeneous, marked by intermittent patches 
of enhanced dissipation and magnetic reconnection in current sheets.  
This small-scale dissipation and reconnection can heat the plasma 
and reshape the large-scale magnetic topology.  Other factors
that can influence the coupling between large and small scales
include magnetic helicity, cross helicity, dynamic alignment,
and the suppression of small-scale turbulence by large-scale 
magnetic flux (\S\ref{sec:backscatter}).  Most of these effects 
are neglected by current SGS models, which often assume some degree 
of isotropy and homogeneity to make paramphenomenaeterizations more 
tractable.  Further investigation of these issues and 
associated SGS model development is sorely needed.   Help in
dealing with inhomogeneity and intermittency may also come
from adaptive mesh refinement (AMR), which was auspiciously
addressed at the GTP Workshop by O.\ Vasilyev \citep{deste13}.

Yet, the challenges in modeling SGS physics do not stop there.
In most plasma flows (particularly those with low density), the
MHD equations cease to be valid on the smallest scales, giving 
rise to kinetic effects that lie outside the scope of ideal
or even resistive MHD.  Such effects may regulate the
dissipation of energy and magnetic helicity, the associated
plasma heating, and the restructuring of the magnetic topology 
through magnetic reconnection.  Kinetic effects also introduce 
new phenomena that may influence large-scale dynamics, including 
non-thermal
particle acceleration and anisotropic heat conduction and 
viscosity.  A promising path forward is to couple MHD models
to kinetic or hybrid codes that capture some of the relevant
kinetic effects (\S\ref{sec:kinetic}, \ref{sec:hybrid}).
But this will not be easy; it is a formidable theoretical 
and computational challenge.

As mentioned several times in this review, a promising way to validate LES models in MHD and to guide their development is to compare them with higher-resolution DNS or ILES.  Preliminary results from astrophysical application have generally been promising \citep{grete15,meheu15} but more work is needed.  Kinetic and hybrid simulations can also be used to motivate and assess SGS models, particularly for models that are not purely dissipative.  We expect to see much progress on these fronts in the next 5-10 years.

There is no oracle or omen to tell you whether your application
requires sophisticated SGS modeling or if ILES is sufficient.  This 
judgement must be made on a case-by-case basis grounded on a thorough 
understanding of the underlying physics and indeed, it is still being
assessed even for the relatively well-established problems surveyed
in \S\ref{sec:applications}.  Though there are robust features 
of MHD turbulence that can be exploited in SGS models, many 
aspects of the SGS physics are likely not universal.  Yet, 
if prudence is followed when designing numerical models and
interpreting the results, current applications do give us 
confidence in the central premise of LES (\S\ref{sec:intro}), 
namely that real heliophysical and astrophysical systems can 
be meaningfully modeled when only a fraction of the dynamically
active scales are explicitly resolved.  LES of MHD turbulence 
is a still a nascent field, brimming with challenges...and 
opportunities.

%========================================================================

\begin{acknowledgements}
We thank the reviewer Wolfram Schmidt and also Hideyuki Hotta for constructive comments that have improved the 
content and presentation of the paper.  
The work of A.\ Petrosyan was supported by the Russian Foundation for Basic Research (14-29-06065) and
Program \#9 of the Russian Academy of Science Presidium ``Experimental and Theoretical Studies of
Solar System objects and stellar planetary systems''.  A.\ Brandenburg gratefully acknowledges 
support from the Swedish Research Council grants
No.\ 621-2011-5076 and 2012-5797 and the Research Council of Norway
under the FRINATEK grant No.~231444.
NCAR is sponsored by the National Science Foundation.
\end{acknowledgements}

%\bibliographystyle{aps-nameyear}
%\bibliography{paper} 
%\nocite{*}

\end{document}